\documentclass[review]{elsarticle}

\usepackage{lineno,hyperref}
\modulolinenumbers[5]

\journal{Information Sciences}

\usepackage{algorithm}
\usepackage{algorithmicx}
\usepackage{algpseudocode}
\usepackage[utf8x]{inputenc}
\usepackage{graphicx}
\usepackage{multicol}
\usepackage{multirow}
\usepackage{xspace}
\usepackage[usenames,dvipsnames]{xcolor}
\usepackage{amsmath}
\usepackage{caption}
\usepackage{subcaption}

\renewcommand*{\k}[0]{\ensuremath{k}\xspace}
\newcommand{\ktree}{\ensuremath{\k^2}-tree\xspace}
\newcommand{\iktree}{I\ensuremath{\k^2}-tree\xspace}
\newcommand{\ktreap}{\ensuremath{\k^2}-treap\xspace}

\newcommand{\ktreapu}{\ensuremath{\k^2}-treap$^U$\xspace}
\newcommand{\ktreapue}{\ensuremath{\k^2}-treap$^{UoE}$\xspace}
\newcommand{\dktree}{d\ensuremath{\k^2}-tree\xspace}
\newcommand{\ktrees}{\ensuremath{\k^2}-trees\xspace}

\newcommand{\kTresTree}{\ensuremath{\k^3}-tree\xspace}

\newcommand{\kntree}{\ensuremath{\k^n}-tree\xspace}
\newcommand{\kones}[0]{\ensuremath{\k^2}-ones\xspace}

\newcommand{\kk}{\ensuremath{\k^2}\xspace}

\newcommand{\rank}{rank\xspace}

\newcommand{\ZEROS}[0]{0s\xspace}

\newcommand{\ONES}[0]{1s\xspace}
\newcommand{\kdouble}[0]{\ensuremath{\k^2}-ones\ensuremath{^{\mathrm{2bits-naive}}}\xspace}
\newcommand{\ksus}[0]{\ensuremath{\k^2}-ones\ensuremath{^{\mathrm{2bits}}}\xspace}
\newcommand{\kdf}[0]{\ensuremath{\k^2}-ones\ensuremath{^{\mathrm{df}}}\xspace}
\newcommand{\klevel}[0]{\ensuremath{\k^2}-ones\ensuremath{^{\mathrm{1\textnormal{-}5bits}}}\xspace}

\newcommand{\kkind}{M\kones}
\newcommand{\kkacum}{CM\kones}
\newcommand{\ikones}{I\kones}
\newcommand*{\koct}[0]{\ensuremath{\k^3}-tree\xspace}

\newcommand{\koctindex}{\koct}

\newcommand{\kt}[0]{\ensuremath{k}\xspace}

\begin{document}
\begin{frontmatter}

\title{Extending General Compact Querieable Representations to GIS Applications\tnoteref{acks}}
\tnotetext[acks]{
Partially funded by: IMFD; Xunta de Galicia/FEDER-UE grants CSI:ED431G/01 and GRC:ED431C 2017/58; Xunta de Galicia/FEDER-UE, ConectaPeme grant GEMA: IN852A 2018/14; 
Xunta de Galicia/GAIN grant Innovapeme: IN848D-2017-2350417; by MINECO-AEI/FEDER-UE grants
Flatcity: TIN2016-77158-C4-3-R, Datos 4.0: TIN2016-78011-C4-1-R and ETOME-RDFD3: TIN2015-69951-R; MICINN-AEI/FEDER-UE grants
Steps: RTC-2017-5908-7 and BIZDEVOPS: RTI2018-098309-B-C32; and EU H2020 MSCA RISE BIRDS: 690941.
An early partial version of this article appeared in {\em Proc
SPIRE'13} \cite{BDN13}. Some parts of this work also appeared in G. de Bernardo's PhD thesis~\cite{Guillermo14}.}

\author[udc]{Nieves R. Brisaboa}
\ead{brisaboa@udc.es}

\author[udc]{Ana Cerdeira-Pena}
\ead{acerdeira@udc.es}

\author[udc]{Guillermo de Bernardo\corref{mycorrespondingauthor}}
\cortext[mycorrespondingauthor]{Corresponding author}
\ead{gdebernardo@udc.es}

\author[uchile]{Gonzalo Navarro}
\ead{gnavarro@dcc.uchile.cl}

\author[udc]{\'Oscar Pedreira}
\ead{opedreira@udc.es}

\address[udc]{Universidade da Coru\~na, Centro de investigaci\'on CITIC, Databases Lab., Spain.}
\address[uchile]{University of Chile, Millenium Institute for Foundational Research on Data (IMFD), Department of Computer Science, Chile.}

\begin{abstract}
The raster model is commonly used for the representation of images
in many domains, and is especially useful in Geographic Information
Systems (GIS) to store information about continuous variables of the space
(elevation, temperature, etc.). Current representations of raster
data are usually designed for external memory or, when stored in
main memory, lack efficient query capabilities. In this paper we
propose compact representations to efficiently store and query
raster datasets in main memory. 
We present different representations for binary raster data, general
raster data and time-evolving raster data. We experimentally compare
our proposals with traditional storage mechanisms such as linear quadtrees or
compressed GeoTIFF files. Results show that our structures are up to 10 times
smaller than classical linear quadtrees, and even comparable in space to
non-querieable representations of raster data, while efficiently answering a
number of typical queries.
\end{abstract}


\end{frontmatter}


\section{Introduction} \label{section:Introduction}

The \emph{raster model} is a logical model widely used for the
representation of data in Geographic Information Systems
(GIS)~\cite{Worboys,SpatialGIS} and for the storage of images in
general. It is mainly used in GIS to store information of
continuous variables, that cover the whole space and for which a specific
value at each point in space may exist. Essentially the raster model represents
this information as matrices of values. A matrix is built by
dividing the space into fixed-size cells, so each cell represents the value of
the spatial feature in the corresponding region. Raster image representations
store the value of each pixel in a cell of the matrix.

The raster model is frequently used in GIS to store data 
related to natural geographic phenomena like temperature, wind speed,  rainfall
level, land elevation, atmospheric pressure, etc. Other not nature-related
information, such as land use, is also suitable to be represented by this model.
The alternative model, the \emph{vector model}, usually represents discrete
variables that have well-defined boundaries, using a collection of points and segments.
This is a good fit for the representation of information related to human-made constructions, rivers, boundaries of lakes and
forests, etc., but not for others that cannot be described with a few points and lines.

In this paper, we focus on the efficient representation of raster
data. As stated before, a raster is essentially a matrix, so an
uncompressed raster representation would use much space (for
instance, a raster image with a resolution of just 0.5 km and worldwide coverage
would require a $80,000 \times 40,000$ matrix, or around 13 GB to store an integer
per cell; modern high-resolution raster imagery can reach much higher
spatial resolution, and therefore require much larger storage space).
Because of this, plain raster representations usually have to be stored in secondary memory. Compressed raster
representations exist, but they are mainly designed to reduce
storage, and do not provide efficient access. Most of them are based
on well-known compression techniques such as run-length encoding or
LZW~\cite{lzw}. In these compressed solutions the space requirements
become much smaller, due to the \emph{locality} of raster datasets
(spatial continuity): close cells tend to have similar values.
However, in most of them the full file, or at least large chunks of
the file, must be decompressed even to display a small region of the
space. A well-known technique, called \emph{tiling}~\cite{gis2017}, divides the
raster in smaller, fixed-size tiles and compresses each tile
independently, providing some level of direct access and taking
advantage of the locality of values to improve compression. For
example, the TIFF image format and its extension for geographical
information $GeoTIFF$\footnote{http://trac.osgeo.org/geotiff/}
support this partition into tiles with different compression
techniques including LZW. Still, tiles must be relatively large to
enable compression.

When data collections are stored by a GIS in a compressed format, such as the
ones we describe before, some of the processing
tasks that involve the complete raster can be performed by
simply decompressing the data. However, many operations would benefit from
direct access to regions (e.g., to display a local map), or the
ability to \emph{find} the cells whose value is within some range. A
classic example of this is the visualization of pressure or
temperature bands~\cite{zhang}, where the raster is filtered to display in a
different way the cells according to the range of values to which
they belong. Another example involves retrieving the regions of a
raster with an elevation above a given threshold, to find zones with
snow alert, or below a value, to find regions with risk of floods~\cite{flood}.
Regular compressed raster representations lack the \emph{indexing}
capabilities on the values stored in the raster that would be
required to answer this type of queries. Therefore, these
representations need to traverse the complete raster in order to
return the cells that contain a given value, even when the results
may be restricted to a small subset of the cells.

There are several approaches to provide direct access to values in a
raster dataset. For instance, we can consider the raster as a
3-dimensional matrix and use computational geometry solutions to
answer any query involving spatial ranges or ranges of values by
means of range reporting queries \cite{comp-geom}. However, these
solutions require superlinear space and therefore they are not
suitable to the large datasets involved. Other representations of
raster data that aim at efficient querying are usually based on
quadtrees~\cite{QT}, particularly variations of the linear
quadtree~\cite{linearQT}, a data structure originally devised for secondary
memory. There exist other quadtree representations~\cite{Chang94,CBLQ} that can
work in internal memory, and they are very efficient for processing complete rasters, but they usually lack
query capabilities to access specific regions or cells with specific
values. An extension of the quadtree to 3 dimensions, or
oct-tree~\cite{sametqtot}, could support those queries in a similar way to the
computational geometry solutions. This structure does not require superlinear
space, but does not provide compression either.

Compact data structures have been a very active research topic for the last few
decades. They aim to represent any kind of information (texts, permutations,
trees, graphs, etc.) in compressed space, while supporting query and processing
algorithms that are able to work over the compact representation. This allows
compact data structures to improve the efficiency of classical data structures,
thanks to being stored in upper levels of the memory hierarchy. However,
regarding spatial information, 
and more precisely, raster data representation and querying, most of the previous
work based on compact representations lacks in advanced query support~\cite{Chang94, Lin97}.

A simple solution to store raster data using compact data structures
could be achieved by reading the raster row-wise and storing the
sequence of values. We could use any compressed sequence
representation \cite{GGV03,GMR06,ISAAC10} to return the cells with a
given value (or a range of values \cite{GGV03,CPM12}) efficiently,
but in this kind of approach restricting the search to a spatial
range becomes difficult. Furthermore, these sequence representations
achieve at best the zero-order entropy space of the sequence, and
this is not a significant space reduction in many cases, since it
cannot fully exploit the spatial locality of values in raster data.

In this paper, we propose several compact data structures for
raster data that efficiently support different queries, particularly
those combining spatial indexing (filtering cells in a spatial
window) with filters on values (retrieving cells with a specific
value or in a range of values). We build on existing compact data
structures that represent sets of points in a kind of compressed
linear quadtree, and upgrade them to efficiently store and query
raster data in different forms: simple binary images, general raster
matrices, and even time-evolving raster data. We experimentally test
our proposals to demonstrate their low space requirements and good
performance in these new application domains. Notice that our data
structures can be conceived as a compact representation for any kind
of matrix. Nevertheless, they rely on the locality of values to
achieve compression, so we focus our evaluation on raster data that
displays spatial continuity.

\section{Previous Concepts}
\label{sec-previous}

\subsection{The \ktree}\label{k2tree}
\begin{figure*}[ht]
\begin{center}
   \includegraphics[width=\textwidth]{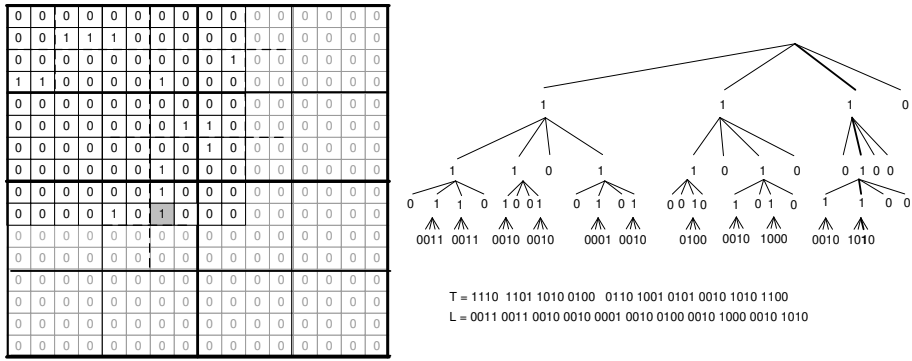}
   \caption{Binary matrix and its \ktree representation, for k=2. The matrix is artificially expanded to the next power of k.}
   \label{fig:k2tree}
\end{center}
\end{figure*}

The \ktree~\cite{BLN14} is a compact data structure for the representation of
sparse binary matrices, that was initially devised to represent the adjacency
matrix of Web graphs, and later applied to compression of social
networks~\cite{Claude11} or RDF databases~\cite{Alvarez15}.  Given a binary matrix of size $n \times n$, the \ktree
conceptually represents it as a \kk-ary tree, for a given \k\footnote{The size of the matrix $n$ is assumed to be a power of
\k. If it is not, the matrix is expanded to the next power of \k filling
the new cells with 0s.}. The root of the conceptual tree corresponds to the
complete matrix. Then, the matrix is partitioned into \kk equal-sized
submatrices of size $\frac{n}{\k} \times \frac{n}{\k}$, and each of them (taken
from left to right and top to bottom) is represented as a child of the root
node. A single bit is associated to each node: a 1 is used
if the submatrix associated to the node contains at least one 1; otherwise, the the bit is set
to 0. The subdivision is applied recursively for each node with value 1, until
we reach a matrix full of 0s or the cells of the original matrix. The conceptual
tree is then stored using two bitmaps: $T$ stores all the bits in the upper
levels of the tree, following a levelwise traversal, and $L$ stores only the
bits in the last level. Figure \ref{fig:k2tree} shows an example of \ktree. 

To navigate the tree a \emph{rank structure} over T is built.
This structure is used to compute the
number of ones in the bitmap up to any position ($rank_1$
operation) in constant time, using sublinear space~\cite{rank}. 
The \ktree exhibits a property that provides simple navigation over the 
conceptual tree using only the bitmaps and the rank structure: given 
a value 1 at any position $pos$ in $T$, its $\kk$ children will start at
position $pos'=rank_1(T, pos) \times \kk$ of $T$. When the last level is reached, 
$pos' > |T|$, so the excess $pos'-|T|$ determines their position in $L$. A \ktree can
answer single cell queries, queries reporting a complete row/column or general range
queries (i.e., retrieve all the 1s in a range) using only
rank operations to traverse the tree, by visiting all the necessary subtrees.

Some improvements have been proposed by the original authors of the
\ktree~\cite{BLN14} to enhance its compression and query efficiency.
For instance, a $hybrid$ \ktree uses different values of \k in the upper and lower levels of decomposition.  
Other techniques use statistical compression of the bitmap $L$.
A dynamic variant of the \ktree, called \dktree, has also
been proposed~\cite{dinamicojournal}. The \dktree is based on a custom 
implementation of dynamic bitmaps for $T$ and $L$. By supporting
update operations over $T$ and $L$, in addition to rank and select operations, the \dktree is able to handle changes in the bits of
the binary matrix, as well as insertion of new rows/columns at the end of the matrix.

\subsection{The \iktree}\label{Ik2tree}
The \emph{Interleaved} \emph{\ktree}~\cite{Alvarez17} (\iktree) 
is a data structure based on the \ktree and devised to deal with RDF triples.
Given a ternary relation $T=\{( x_i,y_j,z_k)\} \subseteq X \times Y \times Z$, the \iktree uses vertical partitioning 
to decompose $T$ into $|Y|$ binary relations $T_j$, one for each different value $y_j \in Y$.
Hence, each adjacency matrix $T_j$ will store the pairs $(x_i,z_k)$
that are related with $y_j$. The dimension $Y$ is called
\emph{partitioning variable}. After this transformation, each of the binary relations could simply be
stored in a separate \ktree, but the \iktree is able to represent all those matrices
simultaneously in the same tree, providing indexing capabilities
also on $Y$. 

\begin{figure*}[ht]
\centering
  \includegraphics[width=0.8\textwidth]{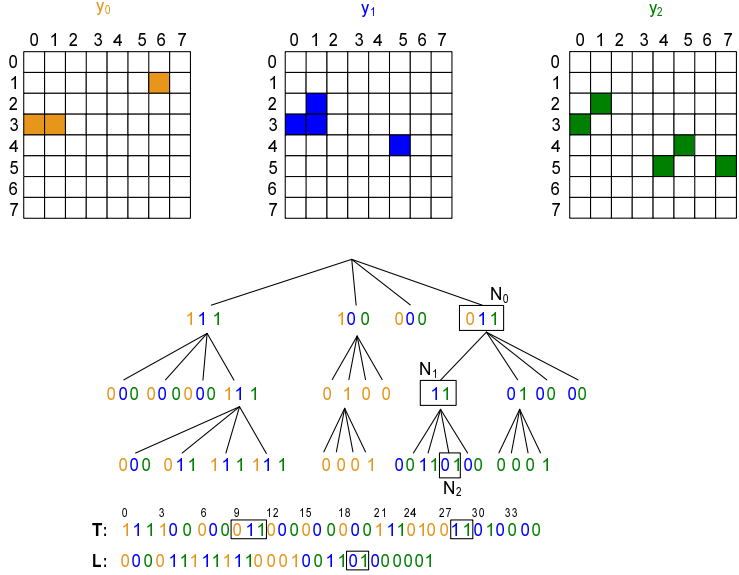}
\caption{Representation of a ternary relation using the \iktree.}
\label{fig:interleaved:structure}
\end{figure*}

Conceptually, building an \iktree is equivalent to building a collection of
\ktrees and merging the equivalent branches of the conceptual trees into a
single tree, where each node will store the bits of all the \ktrees.
This means that the children of the root node will always have $|Y|$ bits, but
nodes at lower levels of the tree have as many bits as 1s exist in their parent
node (i.e., as many bits as trees contain that node). The conceptual tree is
stored in two bitmaps $T$ and $L$, exactly like a \ktree. Figure~\ref{fig:interleaved:structure} displays an example of \iktree, for
$|Y|=3$. Note that the fourth node at the first level of the \iktree ($N_0$)
has 3 bits, one per matrix; its first bit is 0, because the bottom-right
submatrix of matrix $y_0$ is full of 0s, and the second and third bits of $N_0$
are set to 1; therefore, its children have 2 bits each.

The \iktree can be navigated in a similar fashion to \ktrees: at the root
level we have \kk nodes of $|Y|$ bits each; given a node at position $p$, with
$b$ bits, its children are located at position $(rank_1(T,p-1) + c) \times
\kk$ in $T$, where $c=|Y|$ is a fixed correction factor; each children of the
node will have $o$ bits, where $o = \rank_1(T,p+b-1)-\rank_1(T,p-1)$ is the number of bits set to 1 in
the current node. Observe also that, thanks to having the bits for all the $y_i$
together in the same node, it is possible to restrict traversal of the tree to a
specific value $y_j$ in the partitioning dimension. However, pruning the tree
by the $Y$ dimension requires more complex operations than filtering branches on
the other dimensions, so the structure is usually limited to domains where a
partitioning variable of small size can be selected. See~\cite{Alvarez17} for
a further details on the implementation of query operations. in the \iktree.

\subsection{The \ktreap}\label{ktreap}

The \ktreap~\cite{k2treap} is another proposal based on the \ktree, 
and also inspired by the treap~\cite{SA96}. It is specifically designed to
answer range top-\emph{k} queries on multidimensional grids (e.g. OLAP cubes).
Given a matrix and a spatial window inside the matrix, a
range top-\kt query asks for the location of the \kt highest values in the query
window.

Starting with a matrix $M[n \times n]$, where each cell 
can be empty or store a numeric value, the \ktreap
follows a recursive partition of the matrix into \kk submatrices, similar to the \ktree.
The decomposition works as follows: the root of the tree stores the coordinates
of the cell whose weight is the maximum value in the matrix, as well as the cell value. Then, the cell 
is marked as \emph{empty} and removed from
the matrix. Then, the resulting matrix is subdivided into \kk
submatrices and we add the corresponding \kk children nodes to the root of the
tree. The assignment process is repeated for each child, taking the cell with
the maximum value and its coordinates from the corresponding submatrix, and
deleting the value of the cell before continuing. Decomposition eventually stops
when a completely empty submatrix is found or the cells or the original matrix are reached.

The conceptual \ktreap is stored using three elements: $i)$ a
sequence \emph{coords} per level, keeping the coordinates of
the local maxima, and stored as relative offset to the origin of the current
submatrix (note that empty nodes are dismissed, and coordinates are not needed
in the last level of the tree since submatrices are of size 1); $ii)$
the \emph{values} of the local maxima (i.e.
their weights), also differentially encoded with respect to their parent node
and compressed using DACs~\cite{BLN12} (notice that a small array $first$ is also
stored to mark the offset in the array where each level of the tree starts); and
$iii)$ the tree topology, stored like a \ktree with a single bit array, $T$, with \emph{rank} support. This change is necessary since rank operations are also needed in the last level of
the tree in the \ktreap.

\begin{figure*}[!ht]
 \centering
 \includegraphics[width=0.9\textwidth]{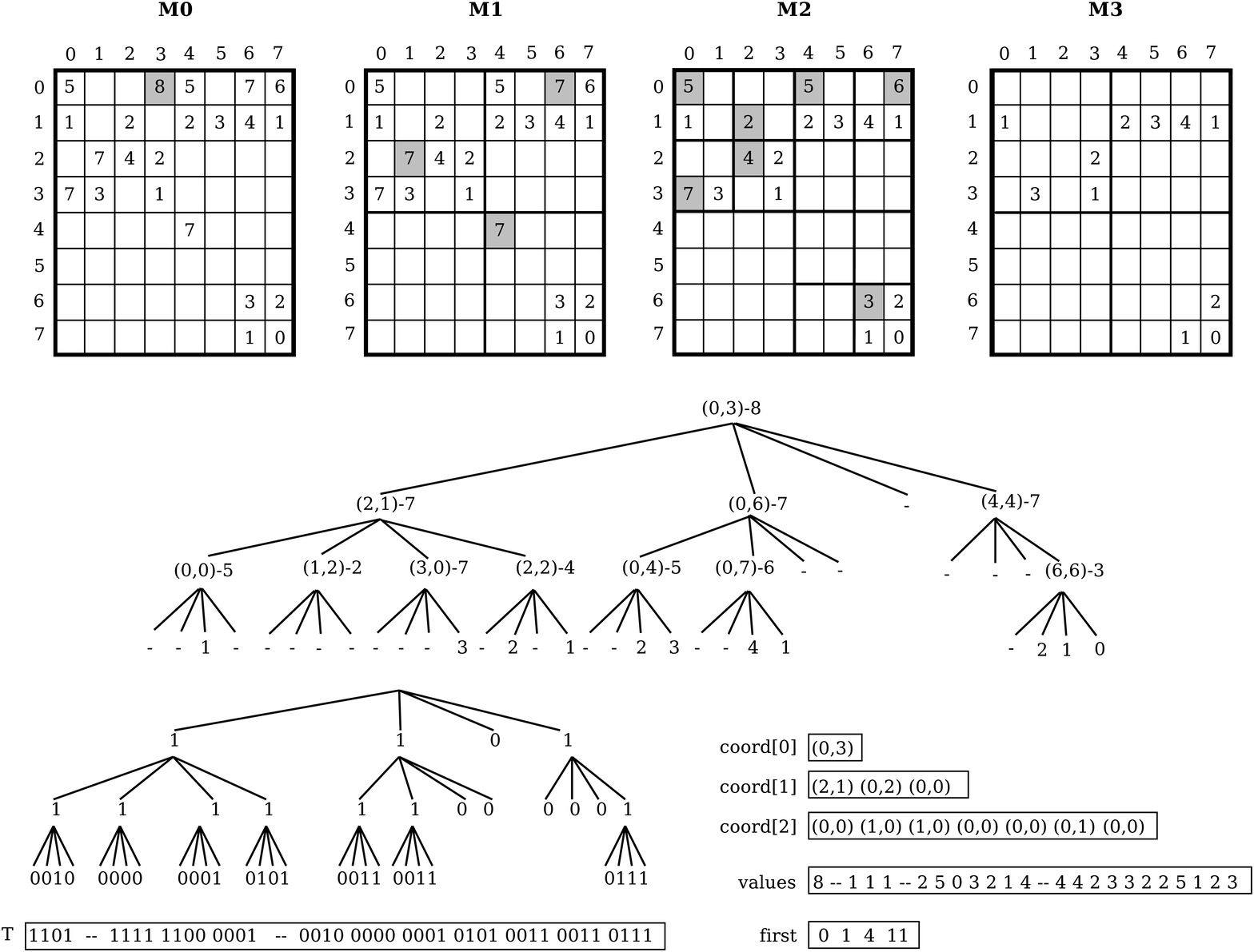}

\caption{Example of \ktreap construction from a matrix and data
structures used to represent it.} \label{fig:k2treap}
\end{figure*}

Figure~\ref{fig:k2treap} shows a \ktreap construction,
for $k=2$. The top of the image shows the state of the matrix
at each level of decomposition, and the cells selected as local maxima at each
level are highlighted, except in the last level where all the cells are
local maxima. Empty submatrices are represented in the tree with the
symbol ``\texttt{-}''.

The \ktreap provides support for cell access, basic range and
top-\emph{k} queries, and also interval queries regarding cell
weights. A detailed description of the structure and navigation
algorithms is presented in \cite{BDBKNS16}.

\section{Representation of binary rasters}
\label{binary-rasters}

Binary images can be considered as the simplest form of raster data. In a binary
image we store a matrix that uses a single bit per cell, to determine whether a
single feature is present or not within the region of the space corresponding
to that cell. Hence, a binary raster is essentially a simplified version of a
general raster, limiting the range of possible values to two. Several GIS
applications make use of this simple technique to represent binary attributes of
the space. Examples of this would be information of events like oil spills,
plagues or cloud cover in their simplest version, as well as simple
rasterized representations of vectorial data.

Due to the simplicity of these binary images, their representation usually
requires specific techniques to achieve the best compression and query
performance. The \ktree, introduced in Section \ref{k2tree}, is an example of
those. However, it was devised to compress Web graphs, so it works well mostly
on binary matrices that are very sparse.
 
In this section we propose a solution, that we call \kones, based on the
\ktree, designed to efficiently compress the kind of binary images that
usually appear in GIS applications. Essentially, our technique is devised to
overcome the limitation of the \ktree to sparse matrices: our technique is
designed to efficiently compress binary matrices with a large percentage of
\ONES, as long as there is some clusterization of the values, which is typical
of most real-world raster data. 

Our proposal is based on the same decomposition of the binary
matrix used by the \ktree, but we recursively divide the matrix until we reach
any uniform region, be it full of \ZEROS or \ONES. This means that in our
\kones we have 3 possible types, or ``colors'', of node, following the usual naming
of quadtrees: \emph{black} and \emph{white} nodes are regions full of ones and
zeros, respectively; the internal nodes, that are regions with ones and zeros,
are \emph{gray}. Note that the main difference of our proposal with a \ktree is
that we are able to represent large regions of ones with a single node, instead
of using a full subtree.

The \kones can efficiently answer cell retrieval queries, as well as row/column
or range queries, simply by performing a top-down traversal of the tree branches
that intersect the region of interest. The only consideration is that when a
black node is found, we need to output all the cells of the region of interest
that fall within the submatrix covered by that node.

\begin{figure*}[ht!]
   \centering
        \includegraphics[width=\textwidth]{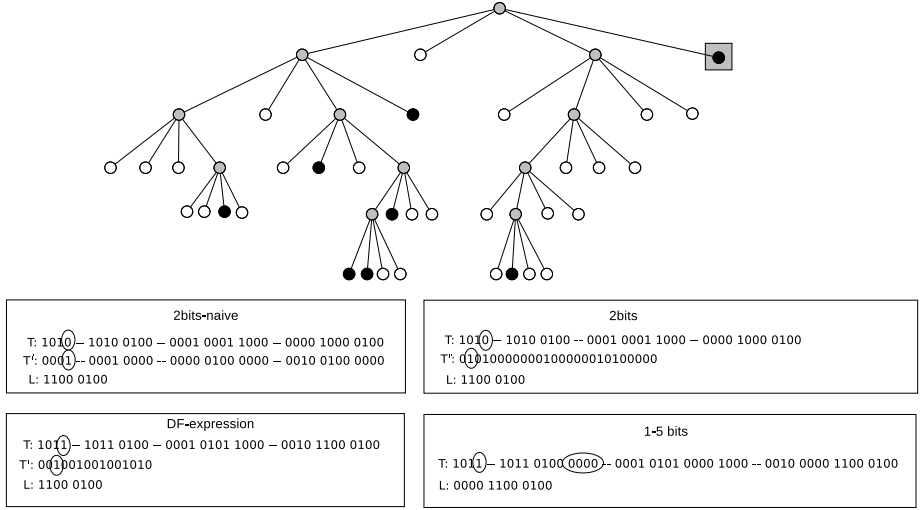}
    \caption{\kones conceptual representation and implementation variants.}
    \label{fig:konesvariants}
\end{figure*}

The goal of the \kones is to be efficiently traversed in a similar
manner to the \ktree, using just rank operations. To achieve this
purpose, we devised a small set of implementation alternatives that store the
conceptual tree in different ways. Figure~\ref{fig:konesvariants} shows all 
our implementation variants for the same conceptual tree. For each variant we
will describe how the components are built, and how the basic traversal
operations are implemented, since query algorithms are based on the same
conceptual traversal of the tree in all cases.

\subsection{Naive 2-bit coding: \kdouble}

A simple representation for the new conceptual tree, where three kinds of
nodes exist, is just to use 2 bits per node instead of one. According to this idea,
we use the following encoding: internal (gray) nodes are encoded with 10; white
nodes with 00; black nodes with 01. Then, the first bit of each node is stored
in a bitmap $T$, and the second bit in a second bitmap $T'$. With this
setup, the bitmap $T$ marks internal nodes with 1 and leaves with 0, and $T'$
stores the type of leaf. Note that this is not necessary in the last level of
the tree, where no internal nodes can exist, so in the last level we use a
bitmap $L$ like in regular \ktrees.

The \kdouble can be efficiently navigated, much like a \ktree, as
follows: given an internal node at position $p$, we can compute the position
where its children start as $p'=rank_1(T,p)\times\kk$, since each 1 in $T$ is an
internal node that yields exactly $\kk$ children. If we find a leaf node
($T[p]=0$), we check $T'[p]$ to determine whether it is a black or white node.
Notice that the rank structure to perform traversal in constant time is only
needed in $T$ but not in $T'$ or $L$.

\subsection{Improved 2-bit encoding: \ksus}

The \kdouble uses 2 bits per node to represent only three possible node types
in the tree. We can use a more space-efficient encoding using just 1 bit
for internal nodes. In this variant, internal nodes are stored with a
bit 1 in $T$, but do not have a second bit in $T'$. Again, the last level is
stored with a single bitmap $L$.

In the \ksus we can compute the children of a node using the same formula of
the previous approach, since $T$ and $L$ are identical. The only difference is
that, when we reach a leaf node at position $p$, the corresponding bit in $T'$
will not be located at position $p$ but at $T'[rank_0(T,p)]$.

\subsection{Navigable DF-expression: \kdf}

Following the same ideas of the previous variants of using two bitmaps $T$ and
$T'$, we propose a variant based on the DF-expression
encoding~\cite{Kawaguchi80}. In this variant, we encode internal nodes with
10, white nodes with 0 and black nodes with 11. We use the same bitmaps
$T$ and $T'$ for the first and second bits of each node, and a single bitmap $L$ in the last level.

The encoding used by the \kdf has been suggested to be more space efficient than
the previous ones, but it is not as efficient in our implementation since it
requires  more complex computations. Particularly, to compute the children of a
gray node we need to count the number of internal nodes up to the current position, as
$\rank_1(T,p) - \rank_1(T', rank_1(T,p))$. This increases complexity and forces
us to add a rank structure not only to $T$ but also to bitmap $T'$ in order to
perform the previous computation. Notice also that, unlike in previous variants,
we now need to check the bitmap $T'$ to know if the current node is internal or
a (black) leaf.

\subsection{An asymmetric approach: \klevel}

Our last proposal aims at storing our conceptual tree, with three types of
nodes, using the same data structure of the original \ktree, and almost
identical encoding. Internal nodes are encoded with 1 and white leafs with 0,
like in the original \ktree. Black leaves are encoded as a small subtree: an
internal node with $\kk$ white leaves as children. We take advantage of
this configuration, that is not possible in a \ktree, to mark black nodes
using $k^2+1$ (typically 5) bits.

The \klevel can be traversed exactly like a \ktree. The only difference is
that, when we are performing traversal, if we reach a node encoded with 1, we
need to check its children: if all of them are white, the current node is
black. In practice, this can be performed when checking the node, or we can
simply traverse it like an internal node in the \ktree, and in the next step
check whether it was indeed an internal or black node.

The \klevel uses a very asymmetric encoding for the nodes, requiring $\kk+1$
bits for black nodes and only 1 bit for white nodes. However, it also has an
interesting property: since it is identical to a \ktree where regions of ones
are encoded using a shorter subtree, this approach will never exceed the space
of the original \ktree.

\subsection{Experimental evaluation}\label{exp:kones}

In this section we compare the \kones with the original \ktree. We focus on
two different types of data, with fundamentally different characteristics:
Web graph datasets, that are very sparse, and raster data, where there can
be a large percentage of ones. Table~\ref{tab:datasetsk2ones} shows the
datasets used. The Web graph datasets\footnote{Provided by the Laboratory for
Web Algorithmics (LAW) at http://law.di.unimi.it/datasets.php} are very sparse datasets (less than 0.005\% of ones). The raster datasets have
been extracted from the Digital Land Model (MDT05) of the Spanish Geographic
Institute\footnote{http://www.cnig.es}. They are high-resolution (cells of $5
\times 5$ meters) elevation rasters. We took several fragments of the overall
dataset, numbered as shown in Table~\ref{tab:datasetsk2ones}. Datasets $mdt\textnormal{-}A$ and $mdt\textnormal{-}B$ are built by combining several adjacent pieces to build
larger rasters. Note that the original datasets store decimal values; we select
a reference value and build binary matrices by selecting all cells with value below the given threshold.

\begin{table}[ht!]
\centering\footnotesize
\begin{tabular}{|l|l|r|r|}\hline
Type & Dataset         &   \#Size   \\
\hline
\multirow{4}{*}{Web graph} & cnr        &   $325,557 \times  325,557$  \\
	&	eu         &   $862,664 \times 862,664$ \\
	&	indochina  &   $7,414,186 \times 7,414,186$  \\
	&	uk         &   $18,520,486 \times 18,520,486$ \\
\hline
\multirow{7}{*}{Raster} &mdt\textnormal{-}200 &   $3,881 \times  5,461$ \\
& mdt\textnormal{-}400 &   $3,921 \times  5,761$ \\
& mdt\textnormal{-}500 &   $4,001 \times  5,841$ \\
& mdt\textnormal{-}600 &   $3,961 \times  5,881$ \\
& mdt\textnormal{-}700 &   $3,841 \times  5,841$ \\
& mdt\textnormal{-}900 &   $3,961 \times  6,041$ \\
& mdt\textnormal{-}A   &   $7,721 \times  11,081$\\
& mdt\textnormal{-}B   &   $48,266 \times 47,050$\\
\hline
\end{tabular}
\caption{Web graphs used to measure the compression of ones.}
\label{tab:datasetsk2ones}
\end{table}

We run all the experiments on an AMD-Phenom-II X4 955@3.2 GHz, with
8GB DDR2 RAM, running Ubuntu 12.04.1. Our
implementations are written in C and compiled with gcc version 4.6.2. with -O9
optimizations.

\subsubsection{Space analysis}

We evaluate our \kones implementation variants comparing them with original
\ktrees. We use for the comparison Web graphs and raster datasets, with a
threshold set to have 50\% of ones. For all the approaches we use a hybrid
version, where $\k=4$ in the first three levels of decomposition and $\k=2$ in
the remaining levels. 

Table~\ref{tab:compressionones} shows the compression achieved by our techniques
and the original \ktree. We highlight the best compression results for each dataset. In the
first four datasets, Web graphs, the \klevel slightly improves the compression
of the original \ktree, thanks to being able to exploit slightly larger clusters of ones that appear in most Web graphs. However, the
sparsity of the datasets makes all of our other variants larger than the
\ktree.
In raster datasets, due to the much higher percentage of ones, the \ktree
becomes much less efficient than our variants: our techniques are roughly 10
times smaller than the \ktree in all the datasets. The \ksus achieves
the best compression results in all of them, but all the variants are relatively
close.

\begin{table}[ht!]
\centering\footnotesize
\begin{tabular}{|l|r|r|r|r|r|}\hline
 Dataset & \ktree &	\kdouble	&	\ksus	& \kdf	& \klevel \\
\hline
cnr & 3.15 & 4.36 & 3.79 & 3.69 & \textbf{3.14} \\
eu & 3.81 & 5.17 & 4.50	& 4.47 & \textbf{3.79} \\
indochina & 2.03 & 2.60	& 2.25& 2.26 & \textbf{1.92} \\
uk & 2.95 & 4.02 & 3.49 & 3.43 & \textbf{2.91} \\
\hline
mdt\textnormal{-}200 &  0.25	& 0.04	& \textbf{0.03}	& \textbf{0.03}	& 0.04  \\
mdt\textnormal{-}400 &  0.22	& 0.02	& \textbf{0.01}	& 0.02	& 0.02 \\
mdt\textnormal{-}500 &  0.23	& 0.03	& \textbf{0.02}	& \textbf{0.02}	& 0.03 \\
mdt\textnormal{-}600 &  0.22	& \textbf{0.01}	& \textbf{0.01}	& \textbf{0.01}	&
\textbf{0.01} \\
mdt\textnormal{-}700 &  0.23	& \textbf{0.02}	& \textbf{0.02}	& \textbf{0.02}	&
\textbf{0.02} \\
mdt\textnormal{-}900 &  0.24	& 0.04	& \textbf{0.03}	& 0.04	& 0.04 \\
\hline
\end{tabular}
\caption{Compression ratio of our techniques vs \ktrees (in bits per one)}
\label{tab:compressionones}
\end{table}

\begin{figure}
\footnotesize \centering
\begin{minipage}{0.50\textwidth}
\centering
\includegraphics[width=\textwidth]{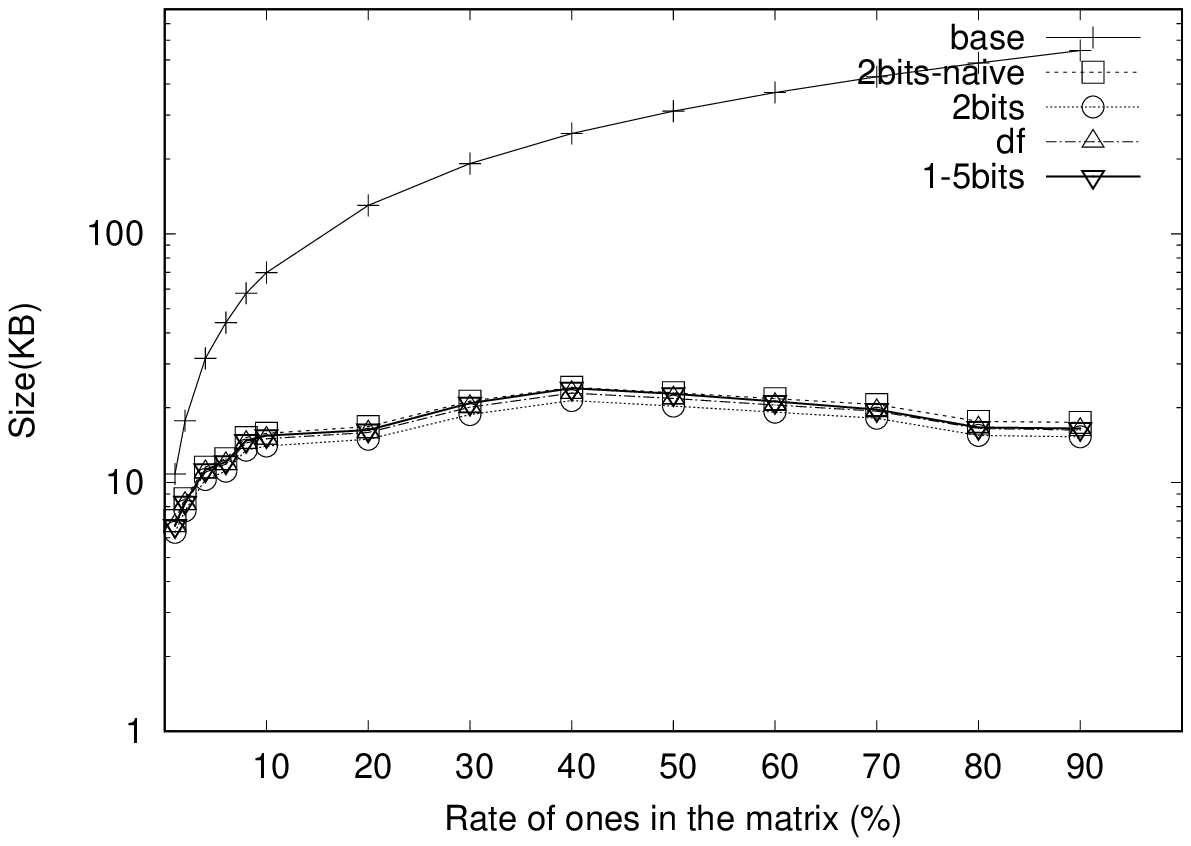}
\end{minipage}\hfill
\begin{minipage}{0.50\textwidth}
\centering
\includegraphics[width=\textwidth]{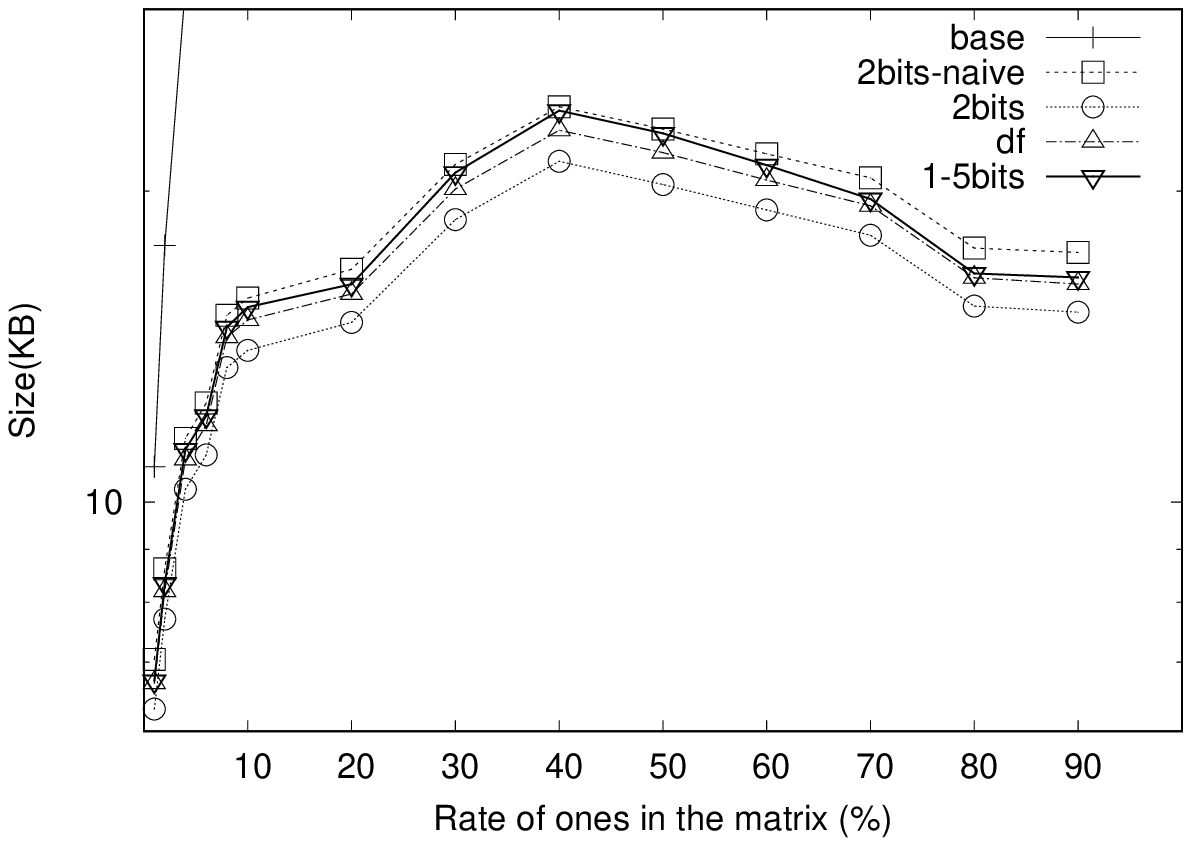}
\end{minipage}
\caption{Space results of \kones variants with different percentage of ones.}
\label{fig:comp-ones-raster-evol}
\end{figure}

To better demonstrate the difference in performance when compared with the
\ktree, we extended the evaluation to binary rasters with varying percentage of
ones. Figure~\ref{fig:comp-ones-raster-evol} (left) displays the compression
obtained for the dataset $mdt\textnormal{-}400$, with thresholds set to get between 1\%
and 90\% of ones. Results show that all the \kones variants are already
smaller than the \ktree baseline with a 1\% of ones in the dataset,
due to the larger size of the clusters of ones. The right-hand plot in
Figure~\ref{fig:comp-ones-raster-evol} focuses on the differences among our
proposals. All of them achieve similar results and evolve almost in parallel,
but the \ksus is the best variant in general and the \kdouble is the worst. The
\klevel, being asymmetric, is slightly worse when the percentage of ones is around 50\%.

\subsubsection{Query times}

In this section we focus on the query performance of the \kones,
particularly compared to that of the \ktree. Specifically, we measure
performance on cell retrieval queries, that involve the traversal of a
single branch of the tree to locate a cell, so they provide a clearer comparison
of the differences in traversal cost among variants.
We perform tests using Web graphs and binary rasters, that are again generated
using a threshold over the original datasets to get binary images with 50\% and 10\% of ones, respectively. To compare the techniques, we measure query times to answer cell retrieval queries, i.e. returning the value of a given cell. We use a set of 10
million random queries for each dataset, and show the average query times in
$\mu$s/query.

\begin{table}[ht!]
\centering\footnotesize
\begin{tabular}{|l|l|r|r|r|r|r|}\hline
Family & Dataset & \ktree &	\kdouble	&	\ksus	& \kdf	& \klevel \\
\hline
&cnr & \textbf{0.46}&0.47&0.52&0.66&0.49 \\
Web & eu & \textbf{0.43}& \textbf{0.43} & 0.48 & 0.61 & 0.45 \\
graphs &indochina & \textbf{0.50} & 0.51 & 0.58 & 0.75 & 0.53 \\
&uk & \textbf{0.58} & 0.60 & 0.66 & 0.89 & 0.61 \\
\hline
&mdt\textnormal{-}200 &  0.54 & \textbf{0.37} & 0.41 &
0.58 & 0.42\\
&mdt\textnormal{-}400 &  0.50 & \textbf{0.29} & 0.33 & 0.43 & 0.34 \\
Raster &mdt\textnormal{-}500 &  0.53 & \textbf{0.33} & 0.37 & 0.51 & 0.38 \\
 (50\%) &mdt\textnormal{-}600 &  0.54 & \textbf{0.27} & 0.30 & 0.40 & 0.31 \\
&mdt\textnormal{-}700 &  0.51 & \textbf{0.29} & 0.32 & 0.43 & 0.33 \\
&mdt\textnormal{-}900 &  0.55 & \textbf{0.36} & 0.41 & 0.56 & 0.41\\
\hline
& mdt\textnormal{-}200 & 0.27 & \textbf{0.24} & 0.27 &
0.34 & 0.26 \\
&mdt\textnormal{-}400 & 0.25 & \textbf{0.22} & 0.24 & 0.30 & 0.24 \\
Raster&mdt\textnormal{-}500 & 0.28 &  \textbf{0.25} & 0.28 & 0.36 & 0.27  \\
(10\%)&mdt\textnormal{-}600 & 0.26 & \textbf{0.23} & 0.25 & 0.32 & 0.25 \\
&mdt\textnormal{-}700 & 0.19 & \textbf{0.15} & 0.17 & 0.20 & 0.17  \\
&mdt\textnormal{-}900 & 0.30 & \textbf{0.28} & 0.31 & 0.41 & 0.31  \\
\hline
\end{tabular}
\caption{Cell retrieval query times (in $\mu$s/query)}
\label{tab:timesones}
\end{table}

Table~\ref{tab:timesones} shows the results for all the datasets, grouped by
family. In Web graphs, the \ktree achieves the best query times, due to
the simpler navigation required. Our encodings obtain higher query times than
original \ktrees. Nevertheless, the overhead of our fastest variant, the
\kdouble, is very low. The \klevel is also very efficient, whereas the \ksus
and especially the \kdf are slower, due to the extra rank operations required.
In the raster datasets, our fastest solutions are always more efficient than the
\ktree, due to the improved access to regions full of ones. Again, the \kdouble
is the fastest variant and the \kdf the slowest. There is also a 
difference in performance depending on the percentage of ones in the dataset:
the \ksus and \klevel are very similar in both cases, but the \klevel is
slightly better when the percentage of ones is lower. Considering that the \ksus achieves the best
compression in all the datasets, we consider it to yield the best space-time
tradeoff overall for any dataset. The \kdouble can offer slightly better
query times sacrificing space, whereas the \klevel can be an alternative
when the number of ones is expected to be relatively low.


\subsection{Comparison with linear quadtrees}
\label{sec-comp-lqt} 

The decomposition of the space in $\kk$ submatrices used in the \kones is a
generalization of the quadrant decomposition used by generic
quadtrees. Hence, our technique can be seen as a compact quadtree representation, since the
conceptual tree we are representing in our variants, for $k=2$, can also be
stored as a classical quadtree. 

The linear quadtree~\cite{linearQT} is a representation devised to work
efficiently from secondary storage. In the linear quadtree, the quadrants are numbered 0-3 from
left to right and top to bottom. Each entry in the matrix (i.e. each 1 in
binary matrices) will be represented by a sequence representing the quadrant
chosen at each decomposition step to reach the corresponding cell. These
sequences, called {\em quadcodes}, can be sorted and stored in a B-Tree in
secondary memory. Cell retrieval queries can be implemented as a simple search
for the corresponding quadcode in the B-Tree.

Our \kones variants are in practice more similar in space to compact quadtree
representations designed for main memory, but those are usually designed
for operations involving the full raster, whereas our techniques still retain
the ability to efficiently access a subregion of the space, something that can
be easily performed with linear quadtrees but not with other compact
representations.
In this section we compare the performance to answer cell retrieval queries of
our techniques against linear quadtree implementations. We implemented an 
in-memory version of the linear quadtree, that uses a B-Tree maintained in main memory.
Additionally, since the linear quadtree is a dynamic data structure that allows
efficient modifications, we perform different comparisons for a \emph{static}
and \emph{dynamic} setup. In the static comparison, we use our \ksus, and
compare it with a linear quadtree that stores quadcodes in an array in main
memory, using binary search to answer queries. In the dynamic comparison, we
use a linear quadtree with a regular B-Tree, fully in
main memory. We use a dynamic version of the \klevel, that is a
straightforward adaptation of the existing \dktree data structure to properly
handle the new semantics for regions of ones. The machine and configuration of
our variants are the same as in Section~\ref{exp:kones}.

\begin{table*}[ht]

\small
\begin{center}
\begin{tabular}{|c|r|r|r|r|}
\hline
\multirow{2}{*}{Dataset}     &   \multicolumn{2}{c|}{Static}   &
\multicolumn{2}{c|}{Dynamic}
\\
\cline{2-5}
            &  \kones  &   Quadtree & \kones    &    Quadtree    \\
\hline
mdt-600$^{50\%}$ &    \textbf{0.02} &   0.25  & \textbf{0.04} &   0.31    \\
mdt-700$^{50\%}$ &    \textbf{0.02}     &   0.17  &  \textbf{0.04}    &   0.23   
\\
mdt-A$^{50\%}$   &    \textbf{0.01} &   0.22   &   \textbf{0.02}   &   0.23   
\\
\hline
cnr         &     \textbf{3.14}    &   41.32  &   \textbf{4.95}     &  
41.46\\
eu          &     \textbf{3.81}   &   49.92 &   \textbf{5.86}      &  
50.07\\
\hline
\end{tabular}
\caption{Compression of \kones and linear quadtrees (in bits per one).}
\label{table-comparison-lqt-space}
\end{center}
\end{table*}

Table~\ref{table-comparison-lqt-space} shows the compression, in bits per one,
achieved by the \kones and the corresponding static and dynamic linear
quadtrees (QT). We only show results for a subset of the collections, since
results are similar among all Web graphs, and among all raster datasets. Results show
that our variants are around 10 times smaller than linear quadtrees in all the
datasets. 

\begin{table*}[ht]
\small
\begin{center}
\begin{tabular}{|c|r|r|r|r|r|r|}
\hline
\multirow{2}{*}{Dataset}     &   \multicolumn{2}{c|}{Static}   &
\multicolumn{2}{c|}{Dynamic}
\\
\cline{2-5}
            &  \kones  &   Quadtree  & \kones    &    Quadtree    \\
\hline
mdt-600$^{50\%}$     &   \textbf{0.25}   &   0.84   &   \textbf{0.56}     &  
0.89    \\
mdt-700$^{50\%}$     &   \textbf{0.28}    &   0.88    &   \textbf{0.61}    &  
0.92    \\
mdt-A$^{50\%}$     &   \textbf{0.26}     &   0.98   &   \textbf{0.71}    &  
1.23    \\
\hline
cnr        &   \textbf{0.77}   &   2.08    &   2.55    &   \textbf{2.28}  
\\
eu          &   \textbf{1.10}    &   2.62    &   3.80    &   \textbf{2.94} 
\\
\hline
\end{tabular}
\caption{Query times of \kones and linear quadtrees (times in $\mu$s/query).}
\label{table-comparison-time}
\end{center}
\end{table*}

Table~\ref{table-comparison-time} displays a comparison of query times. We
measure the average query time over a query set with 1 million random cell
retrieval queries. As shown, our query times are still 2-3 faster than the
linear quadtree in the \emph{static} setup. In the \emph{dynamic} setup, the
overhead required by the dynamic implementation of our structure causes it to
become 2-3 times slower than the static version, so query times become similar
to those of linear quadtrees. Due to this, we are faster than linear quadtrees
in the raster datasets, but slower in Web graphs. We consider the raster datasets to be more
significant to the actual performance of the solutions, since they are designed
for this kind of data, but even the worse query times obtained in Web graphs
are easily compensated by the much better (8x) compression.

\section{Representation of general rasters and spatio-temporal data}
\label{sec-general-evolving}

In this section we introduce solutions based on the \kones that can
handle more complex raster data. Particularly, we focus on the representation of
general raster data and temporal raster data. In general rasters we have a
matrix of non-binary values in which each cell contains a numeric value. Temporal rasters store the
evolution of a raster data along time. We will describe the usual problems for
both kinds of raster data and then introduce our proposals to store them.

In our representations for general raster data we aim at providing
support for queries involving not only the spatial dimension of the dataset, but
also the possible values stored. For instance, the values above a given threshold in an
elevation raster can be selected to yield snow alerts in a given region. Our
solutions are designed to efficiently answer this kind of queries, combining a
spatial constraint with a filter on the possible values, as well as simpler
queries involving constraints only on space or values.

Due to their characteristics, some of the data structures we introduce for
integer rasters can also be adapted to the representation of spatio-temporal data, or
time-evolving regional data. We consider temporal rasters containing the
evolution of a binary raster dataset along time. Hence, we essentially have a
collection of rasters corresponding to the same feature in different time
points. In these datasets, we also have two ways to filter the data: spatial
constraints, to obtain values in a region, and temporal constraints, to obtain
values in a given time interval. We consider the following temporal constraints:
\begin{itemize}
  \item \emph{Time-instant}, or time-slice, queries refer to a single point in
  time.
  \item \emph{Time-interval} queries refer to a time interval. We consider
  three different types of interval: standard queries just return all the
  results found, possibly with multiple occurrences for the same cell;
  \emph{weak} queries will return the set of cells that fulfilled the query
  constraints at any point in the interval (e.g., in a cloud cover raster, find the regions that were covered at any time);
  \emph{strong} queries return the set of cells that fulfilled the constraints
  during the full interval.
\end{itemize}

\subsection{Our proposals}

The proposals we introduce next are \ktree variants, in most cases built
from our \kones implementations. For general rasters, we assume that our input
is a matrix $M$, of size $n \times n$ whose cells contain integer values in the
range $[1,|V|]$. Note that this implies the assumption that the number of
different values is not too large, and raster dataset with floating-point
values can either be rounded or mapped to an integer range. For temporal
rasters, we assume that we have a collection $T$ of binary rasters of the same
size. Most of our proposals can be applied to both cases, with adjusted
algorithms to answer the relevant queries.

\subsubsection{Multiple \kones: \kkind}
\label{sec-ourproposals-indk2tree}

The \kkind uses a collection of \kones to store the original data. If we
see the input matrix $M$ as a collection of binary matrices $M_i$, one for
each possible value, the representation of $M$ is reduced to the
representation of a collection of binary rasters. The \kkind simply stores each
$M_i$ (i.e. the cells with each possible value) in a different \kones $K_i$.

\begin{figure}
\includegraphics[width=\textwidth]{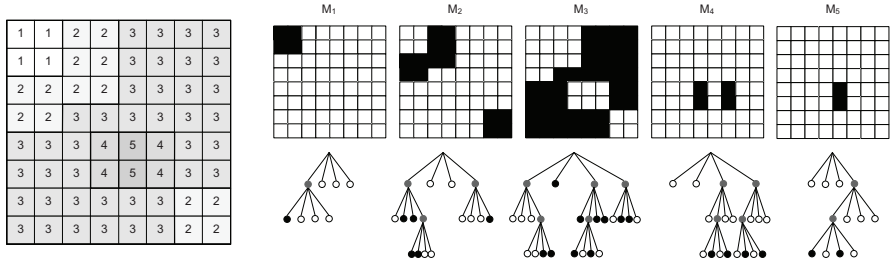}
\caption{Raster matrix and its \kkind representation.}
\label{fig:k2raster-base}
\end{figure}

In this approach, queries involving cells with a given value can be answered
by checking a single \kones. Queries involving a range of values, however,
require checking all the trees in the range, so they become less efficient. The
worst performance, therefore, is expected in queries with no constraints on
values, where all the trees have to be checked.

The same approach can be used for temporal raster data: we use a different
tree per time instant. Time-instant queries are executed on a single tree but
time-interval queries require a synchronized traversal of several trees.
Note that in standard time-interval queries we can just return all the results
querying each tree separately, but for \emph{weak} and \emph{strong} queries we
need to traverse all the trees simultaneously and compute the \emph{or} or
\emph{and} operation of their corresponding bits to filter out branches that do
not fulfill the query semantics.

\subsubsection{Cumulative \kones: \kkacum}
\label{sec-ourproposals-acck2tree}

\begin{figure}
    \centering
    \includegraphics[width=0.95\textwidth]{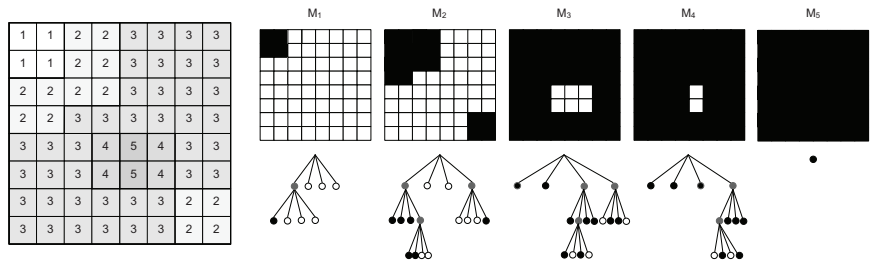}
    \caption{Raster matrix and its \kkacum representation.}
    \label{fig:k2raster-acum}
\end{figure}

Our second proposal, the \kkacum, is based on the same idea of building a tree
per value, but uses a cumulative approach: the first tree will store the cells
with the minimum value; each consecutive tree will store the cells with the next
value, plus all the cells stored in previous trees.
Figure~\ref{fig:k2raster-acum} shows the \kkacum representation, for the same
input matrix of Figure~\ref{fig:k2raster-base}.

In this approach, the trees store a much larger number of ones. However,
taking advantage of the ability of the \kones to store large regions of
ones, the space of the final structure is not expected to increase too much with
respect to the previous approach. In some raster datasets, where values tend to
form concentric curves, the use of cumulative values can even improve
compression by generating larger clusters.

The \kkacum can answer any query
involving a single value, or range of values, using the same strategy: for a range $[\ell, r]$, we compute the results for
value $r$ and subtract those of value $\ell-1$ (in practice, we can traverse
both trees simultaneously to filter out branches as soon as possible). Hence, its
performance is independent of the length of the range. Additionally, it can
answer queries not involving value constraints more efficiently: to find the
value of a single cell, instead of checking every tree, we can use binary search
to look for the leftmost tree that contains the cell.

The \kkacum relies on the fact that the leftmost tree containing a 1 for
the cell yields the actual value of the cell. This approach cannot be used
for time-evolving data, where the same cell can change value several
times.

\subsubsection{\koctindex}
\label{sec-ourproposals-k3tree}

The \koctindex is a straightforward extension of the \ktree to three
dimensions. The conceptual decomposition of a bi-dimensional matrix can be
extended to any number of dimensions, creating $k^n$ submatrices at each step to
build a \kntree. Navigation of the tree is similar, just considering constraints
in the new dimensions and adjusting the formulas to nodes with $k^n$ children.

Our approach uses a \koct to store the complete raster matrix. Particularly,
it will store a 3-dimensional binary matrix, where the third dimension is the
value of the cell. Hence, for each coordinate the only 1 in the third dimension
will correspond to the value of that cell. 

Retrieval algorithms in the \koct are quite simple: to get the value of a cell, we simply traverse 
the conceptual tree looking at all the branches for that $(x,y)$ coordinate; to
find cells with a given value or range of values, we fix the range in the third
dimension and search for all the ones in the corresponding slice of the matrix.

The \koct can also be applied to temporal raster data. Considering the third
dimension as time, we can combine all the raster datasets in a single
3-dimensional matrix. Time-instant and standard time-interval queries are
similar to queries on values. Weak and strong time-interval queries can be
processed as standard queries, filtering out repeated values during or after
traversal.

\subsubsection{\ikones}
\label{sec-ourproposals-intk2tree}

The \iktree has been shown to improve the performance of a collection of
\ktrees in other application domains. Therefore, our next proposal is an
adaptation of the same data structure to work with our \kones. This
just requires adjustments in the data structures
and basic navigation operations similar to those performed in individual
\kones. For instance, using the variant based on the \ksus, a second bitmap $T'$
must be added, and additional operations are defined to check the color of a
node and traverse the tree to reach its children.

\begin{figure}[ht!]
    \centering
    \includegraphics[width=0.9\textwidth]{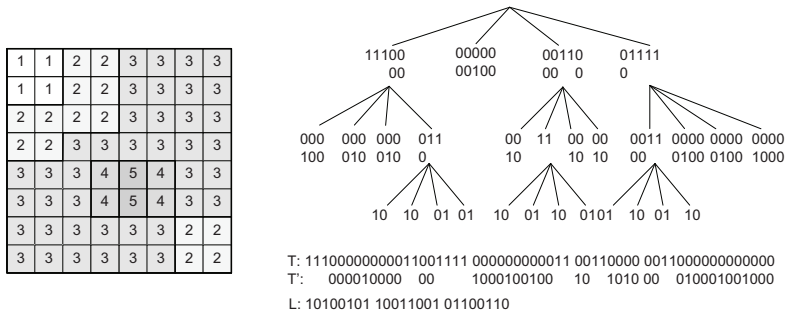}
    \caption{Raster matrix and its \ikones representation}
    \label{fig:interleaved-raster}
\end{figure}

Figure~\ref{fig:interleaved-raster} shows the \ikones, for the same input
matrix used in previous examples. We display the actual bits used by the \ksus
encoding, and the final bitmaps generated. Notice that the bits of each node correspond to the concatenation of the corresponding
bits in the equivalent \kkind representation. 

The \ikones can answer queries involving a single value or a range value using
the same traversal techniques of the original \iktree. Even if navigation is
slower than in individual \kones, making simple queries slower, the ability to
combine all the trees into one provides a much more efficient
way to perform checks in queries involving ranges of values or not involving
value constraints.

The \ikones can also be adapted to temporal raster data. Particularly, most
time-interval queries can be efficiently answered by keeping track of the
corresponding limits of the range for each node: in weak queries, if the current
node contains \emph{at least} a one in our interval, we can confirm the result
immediately; in \emph{strong} queries, if a node has at least a 0 in the
interval, we can discard the result.

\subsection{Experimental Evaluation for General Rasters}
\label{exp:generalrasters}

We test the performance of our proposals using the real elevation rasters
described in Section~\ref{binary-rasters}. Since the values stored are
floating-point values obtained from interpolation, we round the values to a
precision of 1m. 

We compare the compression of our techniques
with a GeoTIFF\footnote{http://trac.osgeo.org/geotiff/} representation of the
same datasets. \emph{tiff$^{~plain}$} simply stores the matrix row-wise,
using a 16-bit integer per cell; \emph{tiff$^{~comp}$} uses the
default compression options: the matrix is partitioned in tiles of size $256
\times 256$, and LZW compression is applied to each tile.

To measure the query efficiency of our proposals, we compare them with GeoTIFF
using the \emph{libtiff} library, version 4.0.3. All time
measurements correspond to CPU time. We consider the following
representative queries:
cell retrieval queries, that ask for the value of a given cell; single-value queries,
that ask for all the cells with a given value; and combined queries, that ask for cells within a spatial region and with values in a given range.

\begin{table}[ht!]
{\footnotesize
\begin{center}
\begin{tabular}{|c|r|r||r|r|r|r|r|r|}
\hline
Dataset     &  \#values   & $H_0$ &   M\kk    &  CM\kk &  $\k^3$  &
$I\kk$    &   tiff$^{~plain}$ &   tiff$^{~comp}$ \\
\hline
mdt-500     &  578             &   5.43    &   2.75    &   2.21    &   1.83    & 
2.53    &   16.01   &   {\bf 1.52}  \\
mdt-700     &  472            &   4.39    &   2.07    &   2.30    &   1.38    &  
1.84    &   16.01   &   {\bf 1.12}  \\
mdt-A       &  978             &   5.86    &   3.24    &   2.83    &   1.94    & 
3.10    &   16.01   &   {\bf 1.52}  \\
mdt-B       &  2,142            &   5.32    &   3.15    &   4.36    &   1.62   
& 3.12    &   16.00   &   {\bf 1.35}  \\
\hline
\end{tabular}
\caption{Compression results in raster datasets (bits/cell).}
\label{table:space}
\end{center}
}
\end{table}

Table~\ref{table:space} shows the compression obtained for different raster
datasets. For each dataset, we show the number of different values
existing in the dataset, as well as the zero-order entropy of the matrix, read
in row order. The best space results are obtained by the compressed TIFF
representation, and the best of our proposals is the \koctindex, that is only
10-20\% larger. Note that \emph{tiff$^{~comp}$} is designed mainly for
compression, and it does not provide support for efficient access. 


\begin{table}[ht!]
{\small
\begin{center}
\begin{tabular}{|c|r|r|r|r|r|r|}
\hline
Dataset     &    M\kk    & CM\kk &   $\k^3$  & $I\kk$   & tiff$^{~plain}$ &  
tiff$^{~comp}$ \\
\hline
mdt-500     &   123.6   &   7.1     &   {\bf 2.2}       &   30.7   &   2.6     &  
491.7   \\
mdt-700     &   65.8    &   6.1     &   {\bf 1.6}       &   27.5    &   2.7    
& 461.9   \\
mdt-A       &   131.9   &   10.2    &   {\bf 2.8}       &   46.2    &   5.2    
& 499.0   \\
mdt-B       &   421.0   &   11.1    &   {\bf 2.9}       &   75.6    &   87.9    &   494.8   \\
\hline
\end{tabular}
\caption{Cell retrieval query times ($\mu$s/query).} \label{table:queryCell}
\end{center}
}
\end{table}

Table~\ref{table:queryCell} shows the results obtained for cell retrieval
queries. Our best approach, the \koctindex, is much faster than the
\emph{tiff$^{~comp}$} variant, and even faster than the plain version (this
is  an artifact due to the nature of the library, that is not designed to access
specific cells and always processes the data in chunks). Among our
techniques, the \kkacum variant is several times slower than the \koctindex,
but still efficient. The \kkind and \ikones variants are much less efficient in
this simple query, in the first case due to the need for a sequential search in
all the trees, and in the second case because of the added complexity of the
structure.

\begin{table}[ht!]
\begin{center}\small
\begin{tabular}{|c|r|r|r|r|r|r|}
\hline
Dataset     &    M\kk    & CM\kk &   $\k^3$  & $I\kk$   & tiff$^{~plain}$ &  
tiff$^{~comp}$ \\
\hline
mdt-500     &   {\bf 3.9}       &   5.8     &   9.4     &   5.9     &   39.5    &   221.4   \\
mdt-700     &   {\bf 3.0}       &   6.0     &   7.3     &   4.5     &   37.5    &   199.5   \\
mdt-A   &   {\bf 8.2}       &   13.6    &   18.9    &   12.7    &   142.6   &   799.0   \\
mdt-B   &   {\bf 110.2}     &   255.1   &   196.6   &   173.5   &   3,838.9 &   19,913.4\\
\hline
\end{tabular}
\caption{Query times for single-value queries (ms/query).}
\label{table:queryValue}
\end{center}
\end{table}

Table~\ref{table:queryValue} displays the query times to retrieve all cells
with a given value. This query demonstrates the indexing capabilities of our
techniques, all of them being much faster than the TIFF-based implementations,
because we can filter results by value while they have to traverse the
complete dataset. The \kkind is the fastest technique, since it has a
specific structure per value. The \kkacum, as expected, is roughly two times
slower. The \ikones is also inefficient, due to the more complex
navigation of the structure. Finally, the \koctindex is now slightly
slower than the other techniques, due to the locality of values: many regions
with values close to the target generate branches in the tree that have to be
checked but will be discarded later.

\begin{table}[ht!]
{\small
\begin{center}
\begin{tabular}{|c|c|c|r|r|r|r|r|r|}
\hline
Dataset     &   Window  &   Range   &   M\kk        & CM\kk     &   \k$^3$  &  
$I\kk$ &    tiff$^{~plain}$  &   tiff$^{~comp}$ \\
            &   size    &   length  &       &   &           &       &               &             \\
\hline
\multirow{4}{*}{mdt-500}        &   \multirow{2}{*}{10} &   10  &   9.0    &  
1.9   &   \textbf{1.8} &   25.9    &   33.0    &   533.0 \\
                                &                       &   50  &   43.1    &  
                                {\bf 2.1}   &   2.6 &   27.9    &   25.0    &  
                                528.0 \\
            \cline{2-9}
                                &   \multirow{2}{*}{50} &   10  &   13.5    &  
                                {\bf 3.4}   &   5.0 &   29.0    &   119.0   &  
                                694.0 \\
                                &                       &   50  &   69.7    &  
                                {\bf 5.9}   &   16.0&   41.2    &   120.0   &  
                                695.0 \\
            \hline
\multirow{4}{*}{mdt-700}        &   \multirow{2}{*}{10} &   10  &   9.6     &  
2.1         & {\bf 1.7} &   24.1&   32.0    &   506.0 \\
                                &                       &   50  &   45.3    &  
                                {\bf 2.1}   &   2.3 &   25.1    &   25.0    &  
                                496.0 \\
            \cline{2-9}
                                &   \multirow{2}{*}{50} &   10  &   13.5    &  
                                {\bf 4.0}   &   4.4 &   29.2    &   123.0   &  
                                649.0 \\
                                &                       &   50  &   68.5    &  
                                {\bf 5.4}   &   13.4&   37.7    &   123.5   &  
                                649.0 \\
            \hline
\multirow{4}{*}{mdt-A}      &   \multirow{2}{*}{10} &   10  &   9.9    &   2.6
&   {\bf 2.0}   &   37.2    &   81.0    &   548.0 \\
                                &                       &   50  &   43.6    &  
                                2.8 &   {\bf 2.6}   &   38.6    &   47.0    &  
                                532.0 \\
            \cline{2-9}
                                &   \multirow{2}{*}{50} &   10  &   13.4    &  
                                {\bf 3.8}   &   4.2 &   39.2    &   228.0   &  
                                703.0 \\
                                &                       &   50  &   62.2    &  
                                {\bf 4.9}   &   11.0&   46.9    &   229.0   &  
                                697.0 \\
            \hline
\multirow{4}{*}{mdt-B}      &   \multirow{2}{*}{10} &   10  &   11.6    &   3.9
&   {\bf 2.3}   &   55.7    &   1,329.0   & 1,265.0\\
                                &                       &   50  &   56.9   &  
                                3.9 &   {\bf 2.5}   &   58.1    &   881.0 &
                                892.0 \\
            \cline{2-9}
                                &   \multirow{2}{*}{50} &   10  &   14.5    &  
                                4.5 &   {\bf 3.2}   &   59.1    &   2,007.0 &
                                1,237.0 \\
                                &                       &   50  &   49.8    &  
                                {\bf 5.5}   &   21.2 &   89.0    &   5,715.0 &  
                                2,038.0   \\
\hline
\end{tabular}
\caption{Query times for combined queries ($\mu$s/query).}
\label{table:queryWindow}
\end{center}
}
\end{table}

Table~\ref{table:queryWindow} shows the query times obtained for combined
queries involving different spatial windows and value ranges. Results confirm that all
our proposals are again faster than the TIFF-based solutions, that are unable
to filter small subsets of data. The \kkacum is now the fastest of our
techniques in most cases, thanks to its ability to efficiently compute the difference between any two values. The \koctindex
also achieves good query times overall, and is the fastest technique in some
of our tests, thanks to its ability to efficiently filter in the 3 dimensions
at the same time. The \kkind is very inefficient, especially with longer
ranges, whereas the \ikones is also inefficient but scales better to longer
ranges.


\subsection{Experimental Evaluation for Temporal Rasters}\label{exp:timevolving}
Next we test the application of our proposals to temporal raster data. We
perform an experimental evaluation on real and synthetic datasets.
CFCA and CFCB contain cloud fractional
cover data\footnote{Obtained from the Satellite Application Facility on Climate Monitoring, at http://www.cmsaf.eu}, covering
the whole world with a resolution of 0.25 degrees. CFCA uses data from years
1982 to 1985, and CFCB data from 2007 to 2009. Our threshold to determine the value of the raster is a cover value above 50\%. 
RegionsA and RegionsB are synthetic datasets created by randomly grouping
circles and altering their borders to build random but generally smooth
and connected regions. Time evolution in these datasets simulates slow movement
and changes/deformations of the original shapes.

The experiments in this section were run in a machine with 4 Intel Xeon E5520
cores at 2.27 GHz and 72 GB of RAM memory, running Ubuntu 9.10. Our code
is compiled with gcc 4.4.1, with -O9 optimizations.

\begin{table}[ht]
\small \centering
\begin{tabular}{|c | c | r | c||r|r|r|r|r|}
\hline
\multirow{2}{*}{Dataset} & \multirow{2}{*}{Size} & \multirow{2}{*}{\#snaps.} &
\% & \multirow{2}{*}{$k^3$} & \multirow{2}{*}{M$k^2$} & \multirow{2}{*}{I$k^2$}   &  
\multicolumn{2}{c|}{Quadcodes}\\
\cline{8-9}
         &&& ones  &   &   &   &   base   &   diff    \\

\hline
CFCA & $720 \times 1440$    & 1,111  & 67.6 &   1.11    &   0.71    &   0.55    &   6.73    &   5.01    \\
CFCB & $720 \times 1440$    & 918   & 58.4 &   1.37    &   0.83    &   0.65    &   7.53    &   5.77    \\
RegionsA & $1000 \times 1000$   & 1,000  & 23.7&   0.04    &   0.09    &   0.06    &   0.64    &   0.16    \\
RegionsB & $1000 \times 1000$   & 1,000  & 24.2&   0.03    &   0.08    &   0.06    &   0.53    &   0.13    \\
\hline
\end{tabular}
\caption{Temporal raster datasets used and compression results (in bits per
one).}
\label{tab:datasets}
\end{table}

%
%

Table \ref{tab:datasets} displays the spatial size, number of time instants
and percentage of ones in each dataset. The remaining columns of the table show
the compression results obtained by our proposals. As a baseline, we show
the space that would be necessary to store the quadcodes of the corresponding raster datasets with two approaches: 
using a separate representation per time instant (\emph{base}); and using a
differential approach where only the changes are stored at each time instant
(\emph{diff}). The latter corresponds to the minimum space that would be
required by a linear quadtree that uses differential encoding, like the OLQ~\cite{BenchSTAM}.
Results show that our techniques are much more space-efficient than
the baseline. The \kkind and the \ikones, that do not take advantage of
similarities between consecutive time instants, achieve the best results in
the CFCA and CFCB datasets. However, in RegionsA and RegionsB the \koct is much
more efficient. This is due to the change rate of the datasets: in the CFC
datasets a large fraction of values change between consecutive time instants,
whereas in our Regions datasets changes are more gradual. Therefore, the \koct
can take advantage of similarities between consecutive time instants in the
latter, but is not able to do it in the former.

In order to confirm the effect of the change rate, we build smaller datasets
taking subsets of 100 snapshots from RegionsA. We build datasets taking every
time instant, every second time instant, and so on, hence representing the
same temporal raster with different time granularity. We also create a new
dataset, built like RegionsA but with $2000 \times 2000$, and generate a
similar group of subsets from it.

\begin{figure}[ht!]
\centering
   \includegraphics[width=\textwidth]{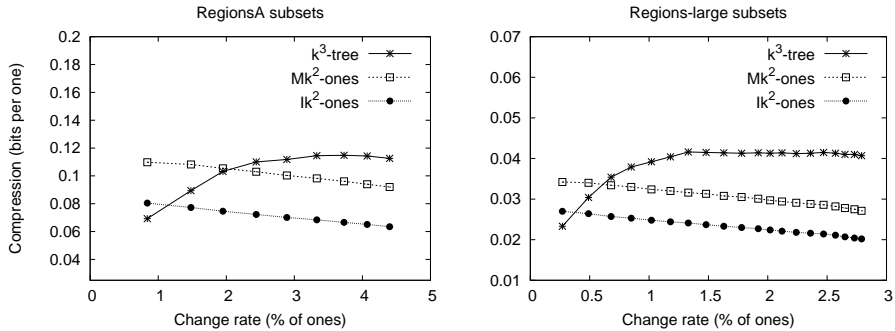}
\caption{Compression results with different change rate} \label{fig:evolComp}
\end{figure}

Figure \ref{fig:evolComp} shows the compression results obtained in the
datasets built from RegionsA (left) and in the datasets built from the
larger raster (right). Each plot displays how the compression obtained by our
structures evolves as the change rate (measured as the percentage of
ones that change on average between consecutive time instants) increases. The
\koct is the most efficient of our proposals for datasets with very small
change rate, but when the number of changes reaches a given threshold,
the implementations that ignore similarity between snapshots (\kkind and
\ikones) become more efficient. Hence, the \koct is the best alternative for
slowly changing datasets, but it is not able to exploit similarities when
changes exceed a relatively slow percentage. Notice that datasets with high
change rate would also be difficult to compress using any other state-of-the-art
techniques based on exploiting this similarity, like OLQs.

We compare the query performance of our proposals using snapshot queries
and time-interval queries. We select different window sizes and time interval
lengths, and build random query sets for each of them. For time-instant queries,
our query sets for each configuration contain 1,000 random queries.
For time-interval queries, we also consider different interval lengths, and
build query sets with 10,000 random queries per window size, interval length
and dataset. In all cases, we measure CPU times, and average the times over a
number of repetitions of the full query set to obtain precise results.

\begin{figure}
    \centering
    \includegraphics[width=\textwidth]{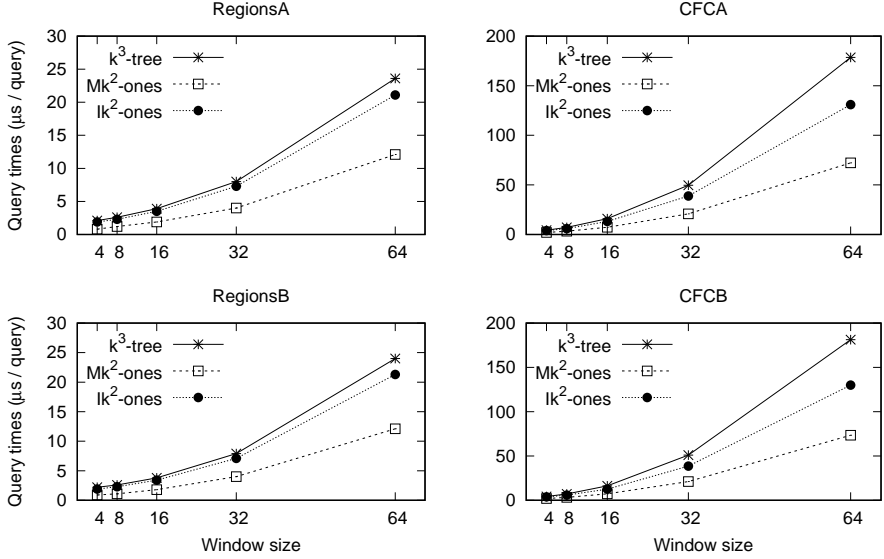}
    \caption{Query times for time-instant windows queries ($\mu$s/query).}
    \label{fig:snapshot}
\end{figure}

Figure~\ref{fig:snapshot} shows the results obtained for all
the datasets in snapshot queries, for different spatial window sizes. Results
are consistent with those in the previous section: the \kkind is the fastest
technique, since it only has to query one tree. The \ikones is around two times
slower, but still faster than the \koct, that must traverse many branches
corresponding to time instants close to the target.

\begin{figure}
    \centering
    \includegraphics[width=\textwidth]{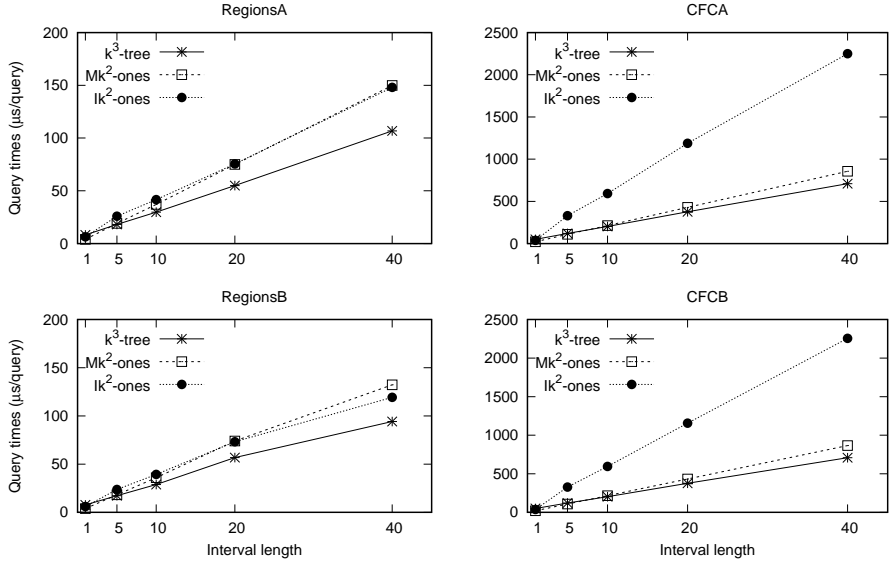}
    \caption{Query times for time-interval queries (window size 32, times in
    $\mu$s/query)}
    \label{fig:interval-w32}
\end{figure}

Figure~\ref{fig:interval-w32} shows the query times for standard time-interval
queries (i.e. queries returning all occurrences for the same cell). We
display results for a representative window of size 32, and interval lengths 1
to 40. The \koct is the most efficient technique for long intervals, whereas
the \ikones is competitive in shorter intervals. Notice that the \kkind is only
the fastest for snapshot queries.

\begin{figure}[ht!]
    \centering
    \includegraphics[width=\textwidth]{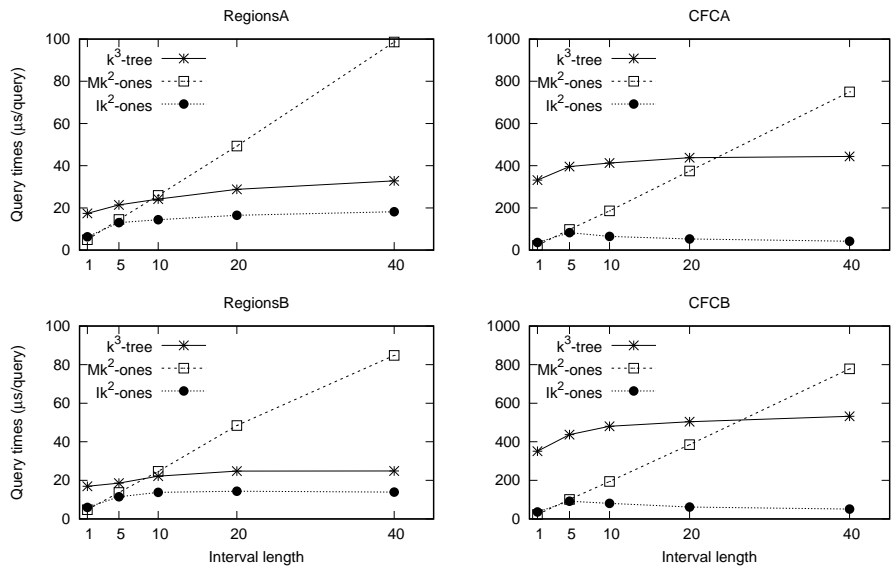}
    \caption{Query times for weak interval queries, for window size 32.}
    \label{fig:weak-w32}
\end{figure}

\begin{figure}[ht!]
    \centering
    \includegraphics[width=\textwidth]{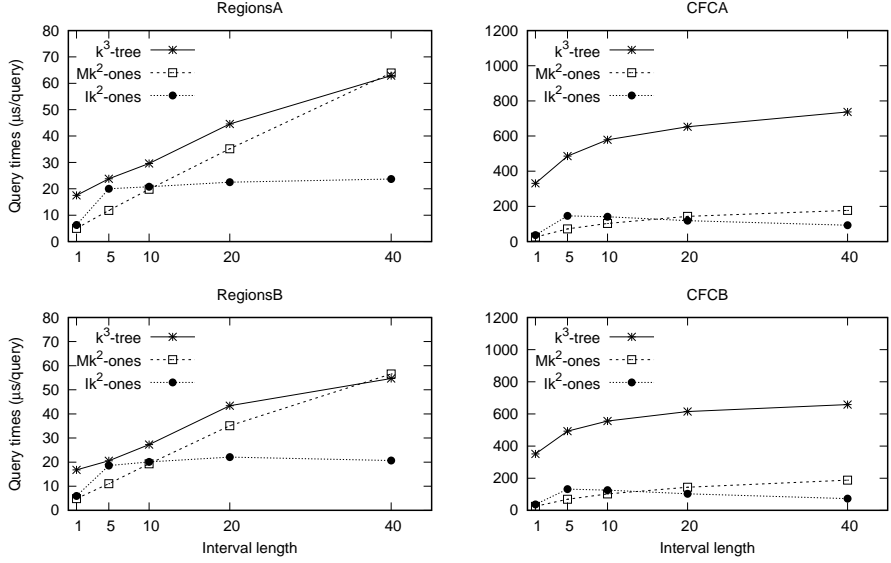}
    \caption{Query times for strong interval queries, for window size 32.}
    \label{fig:strong-w32}
\end{figure}

In addition to standard time-interval queries we also check weak and strong
queries. The results are shown in Figures~\ref{fig:weak-w32} and~\ref{fig:strong-w32} respectively. The evolution of query times is
significantly different for these queries: the \kkind technique still achieves
query times roughly proportional to the length of the interval, since it must
perform a search in all the trees involved. However, the \koct and the \ikones
are much less affected by the interval length. The \ikones obtains similar times
for any interval length, and is the best solution in general in this case,
since it has the ability to efficiently check any time interval at any node of
the conceptual tree. The \koct, on the other hand, cannot improve the query
times of the standard query algorithm, being forced to check all the branches
and then removing duplicates, so it becomes much slower than the \ikones. In
strong interval queries, in which many search branches could be potentially
filtered checking the intervals, the \koct is the slowest technique in general,
especially in the CFC datasets, due to their higher change rate.

\section{Top-\kt range queries in raster data}
\label{sec-topk}

In this section we describe how to apply the same ideas devised in previous sections to
obtain structures that solve top-\kt range queries, i.e. given a spatial window, queries that retrieve the cells with maximum
values inside it. The \ktreap, introduced in Section~\ref{ktreap}, is able to answer this kind
of queries in general matrices. We introduce next two variants that extend the original \ktreap to 
efficiently handle raster matrices where values are highly clustered. Then, we compare our proposals with a naive technique
based on the \kkind, that simply searches for cells in the tree corresponding to the maximum value, and keeps
searching in consecutive trees until the desired number of results is obtained.

\subsection{\ktreap variants}

\begin{figure}[ht!]
    \centering
    \includegraphics[width=\textwidth]{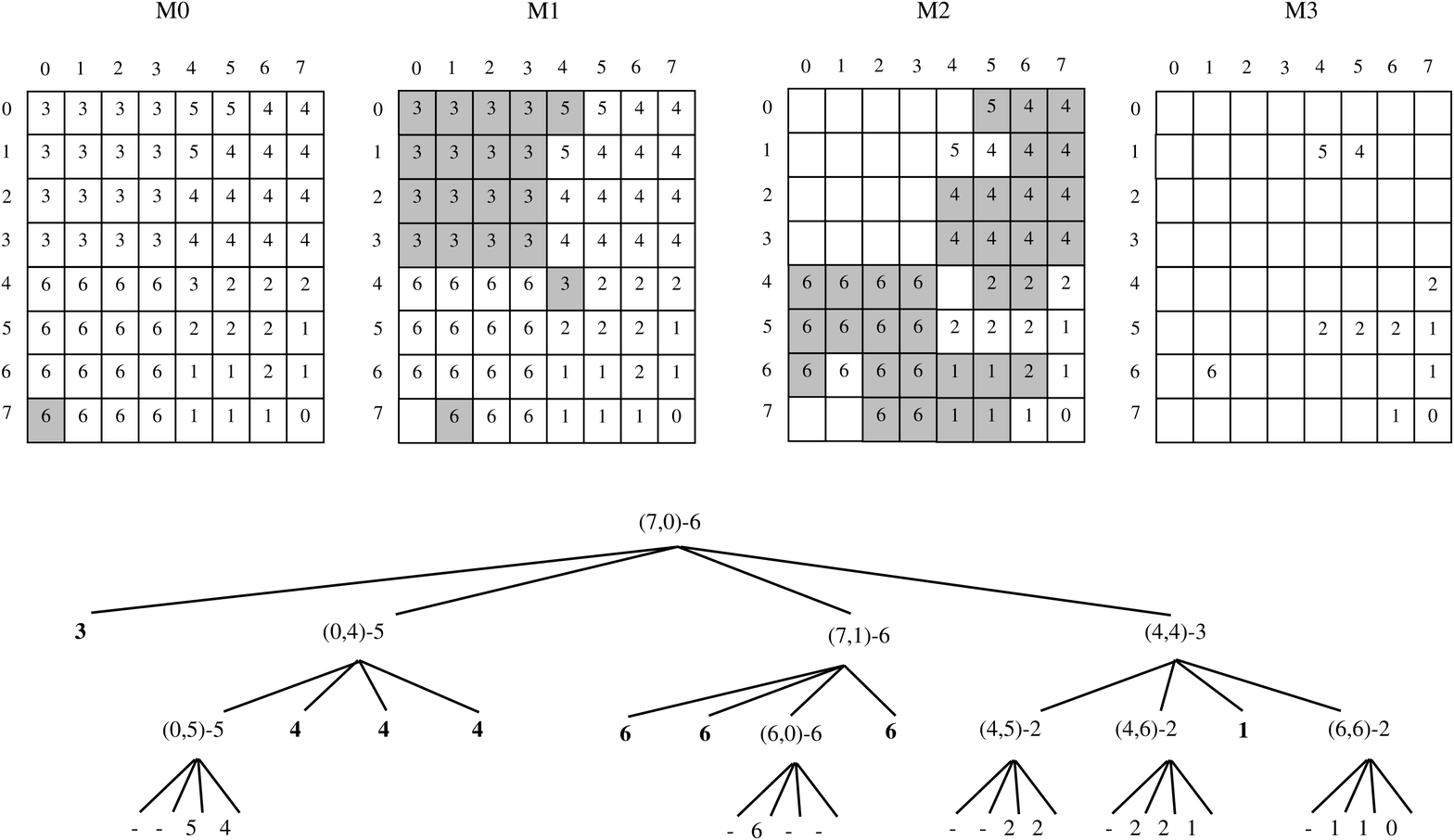}
    \caption{Conceptual representation of the \ktreapu.}
    \label{fig:k2treapCompleto}
\end{figure}

Our first variant, called \emph{\ktreap-uniform} (\ktreapu), is built in a
similar manner to the original \ktreap. Yet, like in our \kones, the decomposition of the matrix stops whenever a ``uniform'' submatrix is found.
This can happen when an empty region is identified or when the same value is shared
by all its cells. Figure~\ref{fig:k2treapCompleto} shows an example of this tree decomposition. 
Matrices $M0$ to $M3$ display the consecutive steps of the \ktreapu
construction, where the \emph{top} cells (cells with the maximum value) for each step are
highlighted. Observe that any dataset can be represented in a more compact way if similar values
are present on many of its submatrices. Notice also that in uniform nodes all the cells in the 
submatrix share the same values, so we do not have to keep the coordinates of the cell with the maximum.

After these changes in the conceptual tree, we use a \ksus to store the tree shape.
Uniform nodes are marked as black nodes were in a binary raster, and empty nodes as white nodes.
Using the same techniques explained for binary matrices, we can easily 
check whether a node is empty or uniform. In empty nodes we stop traversal, and in uniform nodes we can immediately output all the
cells in the submatrix with the same value. The actual representation uses,
in addition to the \ksus, the arrays $coords$, $values$ and $first$, that work
essentially like in the original \ktreap.

Only minor adjustments are required to traverse the conceptual tree in our
variant. Unlike node values, which are kept for all the nodes in the \ktreapu,
coordinates are just stored for non-leaf nodes. Therefore, we can use the formula $\rank_1(T,p) - first[\ell]$ to get the offset in the list of coordinates
corresponding to the current position $p$ and level $\ell$ in the tree. To
compute the offset of the node in the list of values, we also have to consider uniform nodes
(marked with a 1 in $T'$) in our formula: $\rank_1(T,p) + \rank_1\left(T',
\rank_0(T,p)\right)$ (i.e., the number of internal nodes and uniform nodes
that exist up to the current position, respectively).

\begin{figure}[ht!]
    \centering
    \includegraphics[width=\textwidth]{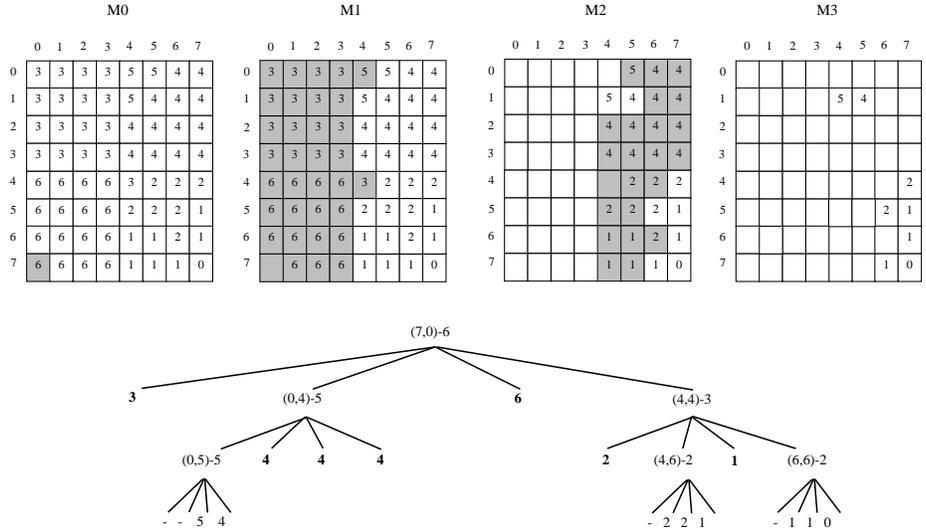}
    \caption{Conceptual representation of the \ktreapue.}
    \label{fig:k2treapUniforme}
\end{figure}

Our second proposal, called \emph{\ktreap-uniform-or-empty} (\ktreapue), tries
to improve compression even more, at the expense of increasing query times. This approach slightly differs from the previous one. 
Here, we stop decomposition at any node as long as
all the values in the corresponding submatrix are equal (even if some cells have a value
and others are empty). For instance, in Figure~\ref{fig:k2treapCompleto}, the
bottom-left quadrant in $M1$ becomes uniform with this new definition. This
variant essentially builds the same \ktreap representation, but taking into account that these regions are now also 
considered as uniform. Figure~\ref{fig:k2treapUniforme} depicts an example of this new approach.

This proposal will cut many branches earlier during the construction of the tree. Even so, it has a drawback:
since we cannot tell apart uniform and empty regions easily, some results may be emitted more than once. 
For instance, if cell $(7,0)$ had the maximum value in the matrix, it will be
emitted at the root of the tree.
But when traversing the bottom-left quadrant, if we identify that region as
``uniform'', it may be emitted a second time.
Hence, to solve top-$\kt$ queries we use an additional data structure to keep track of already emitted results 
(any binary search tree or hash table suffices for this purpose). The additional overhead may become significant
in space and/or time for large \kt, providing a space/time tradeoff between this
proposal and the \ktreapu.

\subsection{Experimental evaluation}

To test the query efficiency of our proposals, we compare them with 
the  \kkind representation of raster data. Notice that, despite its
simplicity, the \kkind can efficiently answer top-\kt
queries by querying the individual trees, starting from the one corresponding to
the highest value, so it should be relatively efficient for this kind of
queries.

Table~\ref{table:raster:spacetreap} shows the compression results obtained by
our \ktreap variants and the \kkind for different raster datasets. Our first
variant, the \ktreapu, is larger than the \kkind, but the \ktreapue achieves
better compression. Both variants obtain reasonable results in terms of space,
at least comparable to the solutions described for general raster data, so they
are a viable alternative if top-\kt queries are relevant.

\begin{table}[ht]
\begin{center}
\begin{tabular}{|c|r|r|r|}
\hline
Dataset     & \kkind & \ktreapu & \ktreapue\\
\hline
mdt-500     &   \textbf{2.75}    &   3.18    &   2.87    \\
mdt-700     &   2.07    &   2.19    &   \textbf{1.98}    \\
mdt-A       &   3.24    &   3.51    &   \textbf{3.20}    \\
\hline
\end{tabular}
\caption{Compression results obtained for \ktree data structures and \kkind (in
bits/cell).} \label{table:raster:spacetreap}
\end{center}
\end{table}


Table~\ref{table:raster:topktreap} shows the query times for top-\kt
queries obtained by all the tested data structures.
For each dataset several window sizes and values of \kt are tested, by
generating sets of random square windows within the bounds of the raster.
Results show that the \ktreapu exhibits a good performance, regardless of the
window size or \kt value, as it is the alternative that achieves the best
results in most of the cases. The \ktreapue, the most compact of the \ktreap
variants, still behaves well for small values of \kt, but when \kt increases the
overhead of keeping track of previous results dominates the query cost.
Also, observe that for larger values
of \kt, the \kkind becomes more competitive with the \ktreap data structures,
since if the query involves many accesses to the tree retrieving cells from one or more \ktrees, it
requires less computation than extracting values one by one from the \ktreap.


\begin{table}[ht!]
{\small
\begin{center}
\begin{tabular}{|c|r|r|r|r|r|}
\hline
Dataset    &   Window &   \kt    & \kkind   & \ktreapu & \ktreapue\\
\hline
\multirow{9}{*}{mdt-500}    & \multirow{3}{*}{100}  &   10      &   164.2 &  
\textbf{15.3} &  15.8   \\
                            &                       &   100     &   178.3 &  
                            \textbf{40.8} &  49.8   \\
                            &                       &   1,000   &   279.8 &  
                            \textbf{233.7} & 385.5 \\
                            \cline{2-6}
                            & \multirow{3}{*}{500}  &   10      &   131.3 &  
                            \textbf{15.8} &16.0 \\
                            &                       &   100     &   143.5 &  
                            \textbf{41.8} & 50.5 \\
                            &                       &   1,000   &  
                            \textbf{217.0} & 230.0 & 372.5 \\
                            \cline{2-6}
                            & \multirow{3}{*}{1000} &   10      &   125.5 &  
                            \textbf{15.8} &16.3 \\
                            &                       &   100     &   134.5 &  
                            \textbf{41.5} & 51.0 \\
                            &                       &   1,000   &  
                            \textbf{200.0} & 219.8 & 358.0 \\
\hline
\multirow{9}{*}{mdt-700}    & \multirow{3}{*}{100}  &   10      &   357.0 &  
\textbf{12.3} & 12.8 \\
                            &                       &   100     &   381.5 &  
                            \textbf{31.3} &37.3 \\
                            &                       &   1,000   &   455.3 &  
                            \textbf{185.0} & 284.8 \\
                            \cline{2-6}
                            & \multirow{3}{*}{500}  &   10      &   309.0  & 
                            \textbf{15.8} & 16.0 \\
                            &                       &   100     &   346.3 &  
                            \textbf{41.3} & 49.0 \\
                            &                       &   1,000   &   495.8 &  
                            \textbf{244.5} & 383.0 \\
                            \cline{2-6}
                            & \multirow{3}{*}{1000} &   10      &   281.0 &  
                            \textbf{16.8} &17.3 \\
                            &                       &   100     &   318.0 &  
                            \textbf{46.3} &54.0 \\
                            &                       &   1,000   &   514.0  & 
                            \textbf{281.8} & 446.3 \\
\hline
\multirow{9}{*}{mdt-A}      & \multirow{3}{*}{100}  &   10      &   493.6 &  
20.4 & \textbf{19.2} \\
                            &                       &   100     &   478.0 &  
                            \textbf{43.2} &50.8 \\
                            &                       &   1,000   &   665.0  & 
                            \textbf{239.2} & 376.0 \\
                            \cline{2-6}
                            & \multirow{3}{*}{500}  &   10      &   426.8 &  
                            23.2 & \textbf{21.6} \\
                            &                       &   100     &   422.0 &  
                            \textbf{50.4} & 56.0   \\
                            &                       &   1,000   &   581.2 &  
                            \textbf{253.6} & 396.0 \\
                            \cline{2-6}
                            & \multirow{3}{*}{1000} &   10      &   422.4 &  
                            \textbf{22.4} & \textbf{22.4} \\
                            &                       &   100     &   419.2 &  
                            \textbf{52.8} & 60.0 \\
                            &                       &   1,000   &   547.2  & 
                            \textbf{260.0} & 408.0 \\
\hline
\end{tabular}
\caption{Query times for top-\kt queries (times in
$\mu$s/query).} \label{table:raster:topktreap}
\end{center}
}
\end{table}

\section{Conclusions}
\label{sec-conclusions}

We have presented several compact data structures for the
representation of general raster data with advanced query support.
Our representations store real raster datasets in small space and provide
efficient access not only to regions of the raster, but also advanced query
capabilities, such as selecting cells with a particular value or range of
values, queries that involve spatio-temporal restrictions, or even
top-\kt queries.

Most of the proposals are based on variants of the \ktree. We propose a
representation, called \kones, that enhances the \ktree so that we can efficiently
compress any kind of clustered binary matrix. Building over this, we
propose compact and indexed solutions for different application domains.
Additionally, most of the approaches introduced can be transformed into dynamic
solutions using a dynamic \ktree.

Overall, our proposals obtain good compression results and are able to answer
a variety of interesting queries. In our experiments we show that our proposals are very compact, several
times smaller than state-of-the-art representations based on linear quadtrees, and still
able to store and query large datasets in main memory. We evaluate our
representations for general raster data, showing their relative strengths and
drawbacks: the \kTresTree obtains very good space results, being close to a
compressed GeoTIFF representation, and shows competitive times in most cases, but the 
variant with independent \kones obtains the best time results to retrieve all
the cells with a given value, and the variant with cumulative \kones obtains the
best results in most of the queries involving ranges of values. Nevertheless,
the results of our proposals are clearly better than the
representations based on GeoTIFF images. We also apply some of the proposals to
the representation of time-evolving raster data. Results show again relative
strengths among our proposals: a \koct is the best solution for slowly-changing
datasets, but as soon as the change rate increases the approaches based on
multiple \kones become smaller.
Finally, we also test new proposals to answer top-\kt
queries in raster data. Our experiments confirm the space efficiency
of the \ktreap variants, that are competitive in space with our
other representations of raster data and faster to answer top-\kt
queries.

We show the scalability of our representations to
efficiently represent rasters with several thousands of different
values. Nevertheless, the space efficiency of most of our proposals
will degrade if the number of different values in the raster becomes
too high. An assumption in our proposals
is that the number of different values in the dataset is not
too high. We claim that in many real-world datasets, even though the
values actually stored may have a high precision, that precision does
not add quality or accuracy after a given
threshold: when measuring features such as temperature,
elevation, pressure, etc. the actual measurements may have
high-precision but the interpolation of values, or even the
simple averaging of measurements, distorts the precision of the
measurements, so for many purposes we can safely reduce the precision of the
values significantly without reducing the quality of the dataset.

The preliminary version of this work inspired several
other research lines. In particular, limitations to handling
large ranges of values were recently addressed in follow-up research
\cite{k2raster}, that extends our original work to support
higher-precision datasets. Our representations are preferable when
high-resolution values are not available or not relevant (e.g., in some applications, high-resolution values are just
interpolations), as well as in domains where the number of
different values is small (e.g., land-use rasters). Additionally, we
have extended our proposals to efficiently store and query
time-evolving data, a challenging problem where other solutions are
difficult to apply due to the particularities of spatio-temporal
queries.









\bibliographystyle{elsarticle-harv}
\bibliography{references}
\end{document}